\documentstyle[psfig]{mn}

\def\degr{\hbox{$^\circ$}}
\def\arcmin{\hbox{$^\prime$}}
\def\arcsec{\hbox{$^{\prime\prime}$}}
\def\fd{\hbox{$.\!\!^{\rm d}$}}

\def\farcs{\hbox{$.\!\!^{\prime\prime}$}}
 
\title[The Stellar Composition of the Star Formation Region CMa R1]
  {The Stellar Composition of the Star Formation Region CMa R1 -- 
   II. Spectroscopic and Photometric Observations of 9 Young 
        Stars\thanks{Based on observations collected at the European 
        Southern Observatory, La Silla, Chile}}
\author[H.R.E. Tjin A Djie et al.]
  {H.R.E. Tjin A Djie,$^1$ M.E. van den Ancker,$^2$ P.F.C. Blondel,$^1$ 
   V.S. Shevchenko,$^3$\newauthor 
   O.V. Ezhkova$^3$, D. de Winter$^4$ and K.N. Grankin$^3$\thanks{E-mail
contact: Herman Tjin A Djie (herman@astro.uva.nl)}\\
   $^1$Astronomical Institute ``Anton Pannekoek'', University of Amsterdam,
       Kruislaan 403, 1098 SJ  Amsterdam, The Netherlands\\
   $^2$Harvard-Smithsonian Center for Astrophysics, 60 Garden Street, MS 42, 
       Cambridge  MA 02138, USA\\
   $^3$Astronomical Institute of the Academy of Sciences of Uzbekistan, 
       Astronomicheskaya 33, Tashkent 700052, Uzbekistan\\
   $^4$TNO-TPD, Stieltjesweg 1, P.O. Box 155, 2600 AD  Delft, The Netherlands\\}
\date{Accepted $<$date$>$.
      Received $<$date$>$}
 
\pagerange{\pageref{firstpage}--\pageref{lastpage}}
\pubyear{2001}
 
\begin{document}
\maketitle
\label{firstpage}

\begin{abstract}
    We present new high and low resolution spectroscopic and photometric data 
of nine members
of the young association CMa R1. All the stars have circumstellar dust at some 
distance as could
be expected from their association with reflection nebulosity. Four stars 
(HD~52721, HD~53367,
LkH$\alpha$~220 and LkH$\alpha$~218) show H$\alpha$ emission and we argue that 
they are Herbig
Be stars with discs. Our photometric and spectroscopic observations on
these stars reveal new characteristics of their variability. We present 
first interpretations of the variability of HD~52721, HD~53367 and the two
LkH$\alpha$ stars in terms of a partially eclipsing binary, a magnetic activity
cycle and circumstellar dust variations, respectively.
The remaining 
five stars show no clear indications of H$\alpha$ emission in their spectra, 
although their
spectral types and ages are comparable with those of HD~52721 and HD~53367. 
This indicates that
the presence of a disc around a star in CMa R1 may depend on the environment of 
the star. In
particular we find that all H$\alpha$ emission stars are located at or outside 
the arc-shaped
border of the H\,{\sc ii} region, which suggests that the stars inside the arc 
have lost their
discs through evaporation by UV photons from nearby O stars, or from the nearby 
($<$ 25~pc)
supernova, about 1~Myr ago. 
\end{abstract}

\begin{keywords}
circumstellar matter -- stars: early type -- stars: evolution -- 
stars: pre-main sequence -- stars: variables -- open clusters and associations: 
CMa R1
\end{keywords}

\section{Introduction}
Herbst \& Assousa (1977) and Herbst, Racine \& Warner (1978) found convincing 
evidence for a scenario in which the formation of the CMa R1 association was 
induced by a supernova explosion about 1~Myr ago. 
More recently, Comer\'on, Torra \& G\'omez (1998) argued that the star 
formation was triggered in pre-existing clouds and that it was only 
accelerated by the supernova shock.  

In a preceeding paper (Shevchenko et al. 1999; paper I) we have confirmed 
that most stars of CMa R1 have not been formed at the same occasion. Our 
objective of this paper is to present and
discuss new optical and ultraviolet spectroscopic and photometric data for nine 
of its members.  A tenth object, Z CMa (HD~53179), has also 
been observed, but since the interpretation of these data is more complicated, 
we will present the results separately (paper III). 
Apart from LkH$\alpha$~220 (B5e) and LkH$\alpha$~218 (B7e), which are close to 
ZAMS, the observed stars 
are of spectral type early B and on or above the Main Sequence (MS) with ages 
much larger than 1~Myr (paper I). It therefore seems useful to inspect their 
spectroscopic and photometric features in order to detect common characteristics 
in their atmospheric and circumstellar envelopes.

   While most of the stars have been observed before photometrically and 
spectroscopically
(e.g. Herbst et al. 1978; Finkenzeller \& Jankovics 1984; Hamann \& Persson 
1992),
the present optical spectra have a resolution which is seven times larger than 
in the earlier
observations and for several stars a large amount of photometric observations 
has been made
during the last decades.   
In addition we have surveyed information from the high and low resolution UV 
spectra, obtained
by the {\it International Ultraviolet Explorer (IUE)}. In section 2 we give a 
survey of the
observations. Section 3 describes the results of these and earlier 
observations. The photometric and spectral variability data 
of the four emission line stars are discussed and interpreted in section 4.
Section 5 summarizes the results and in this section we also
make an attempt to understand these results in terms of the location of the 
stars in the CMa R1 association.
\begin{table}
\centering
\caption{Log of spectroscopic observations.}
\tabcolsep0.07cm
\small
\begin{tabular}{@{}lccccc}
\multicolumn{6}{c}{ESO}\\
\hline
Star	      &	Range [\AA]  &  Date	    &  JD+2400000   &  $R$ [m\AA] & $t_{\rm 
exp}$ [min]\\
\hline 
HD 52721      &	4200--7800   &	11 Dec. 92  &   48967.7653  &   1076  &  ~2\\				

 	      &	6522--6602   &	11 Dec. 96  &	50428.2279  &   65.6  &	 15\\	  
	      &	5849--5920   &	13 Dec. 96  &	50430.1797  &   58.8  &	 15\\	  
	      &	3912--3955   &	14 Dec. 96  &	50431.3548  &   39.3  &	 15\\	  
	      &	8466--8572   &	15 Dec. 96  &	50432.2681  &   85.2  &	 20\\	  
HD 53367      &	4200--7800   &	11 Dec. 92  &	48976.7722  & 	1076  &	 ~2\\	  
	      &	6536--6590   &	16 Dec. 94  &	49702.3472  &   65.6  &	 15\\	  
	      &	6540--6590   &	15 Jan. 95  &	49732.2750  &   65.6  &	 20\\	  
	      &	5860--5910   &	16 Dec. 94  &	49702.2125  &   58.8  &	 20\\	  
	      &	5860--5910   &	16 Jan. 95  &	49733.2887  &   58.8  &	 15\\	  
	      &	3912--3955   &	14 Dec. 96  &	50431.3561  &   39.3  &	 15\\	  
	      &	8466--8572   &	15 Dec. 96  &	50432.2354  &   85.2  &	 20\\	  
LkH$\alpha$ 220& 6400--8950  &  ~1 Jan. 94  &   49353.2103  &   1245  &  25\\  
              & 3600--6800   &  ~2 Jan. 94  &	49354.3145  & 	1563  &  15\\
HD 53623      &	5849--5920   &  13 Dec. 96  &	50430.1921  &   58.8  &  20\\	   
	      &	6522--6602   &	12 Dec. 96  &	50429.2371  &   65.6  &	 20\\	  
HD 53755      &	5849--5920   &	13 Dec. 96  &	50430.1921  &   58.8  &	 20\\	  
	      &	6522--6602   &	12 Dec. 96  &	50429.2013  &   65.6  &	 15\\	  
HD 52942      &	5849--5920   &	14 Dec. 96  &	50431.1600  &   58.8  &	 15\\	  
	      &	6522--6602   &	12 Dec. 96  &	50429.2208  &   65.6  &	 20\\	  
HD 53974      &	5849--5920   &	14 Dec. 96  &	50431.1445  &   58.8  &	 20\\	  
	      &	6522--6602   &	12 Dec. 96  &	50429.2139  &   65.6  &	 ~7\\	  
HD 54439      &	5849--5920   &	14 Dec. 96  &	50431.1719  &   58.8  &	 10\\	  
	      &	6522--6602   &	12 Dec. 96  &	50429.1852  &   65.6  &	 20\\	  
              &	5849--5920   &	14 Dec. 96  &	50431.1807  &   58.8  &	 25\\
\hline
\vspace*{-0.3cm}
\\
\multicolumn{6}{c}{SAO}\\
\hline
Star	      &	Range [\AA]  &  Date	    &  JD+2400000   &  $R$ [m\AA]\\ 	
\hline 
HD 52721      &  3700--4900  &	14 Oct. 89  &  47814.	    &  200\\
              &  3700--4900  &	15 Oct. 89  &  47815.	    &  200\\	 
HD 53367      &  3700--4900  &	15 Oct. 89  &  47815.	    &  200\\	
	      &  3700--4900  &	02 Mar. 91  &  48318. 	    &  200\\	
\hline
\end{tabular}
\medskip
\tabcolsep0.15cm
\begin{tabular}{@{}lcccc}
\multicolumn{5}{c}{{\it IUE} high res.}\\
\hline
Star	      &	SWP   &  Date	      &  LW     & Date\\ 	
\hline
HD 52721      &	10592 &	 12 Nov. 1980 &	R09283 	& 12 Nov. 1980\\	
HD 53367      &	38686 &	 27 Apr. 1990 &	P17816	& 27 Apr. 1990\\
HD 53755      &	27974 &	 21 Mar. 1986 &	R02166	& 24 Aug. 1978\\	
HD 53974      &	10580 &	 10 Nov. 1980 &	R09272	& 10 Nov. 1980\\	
\hline
\end{tabular}
\end{table}
\begin{figure*}
\vspace*{0.15cm}
\centerline{\psfig{figure=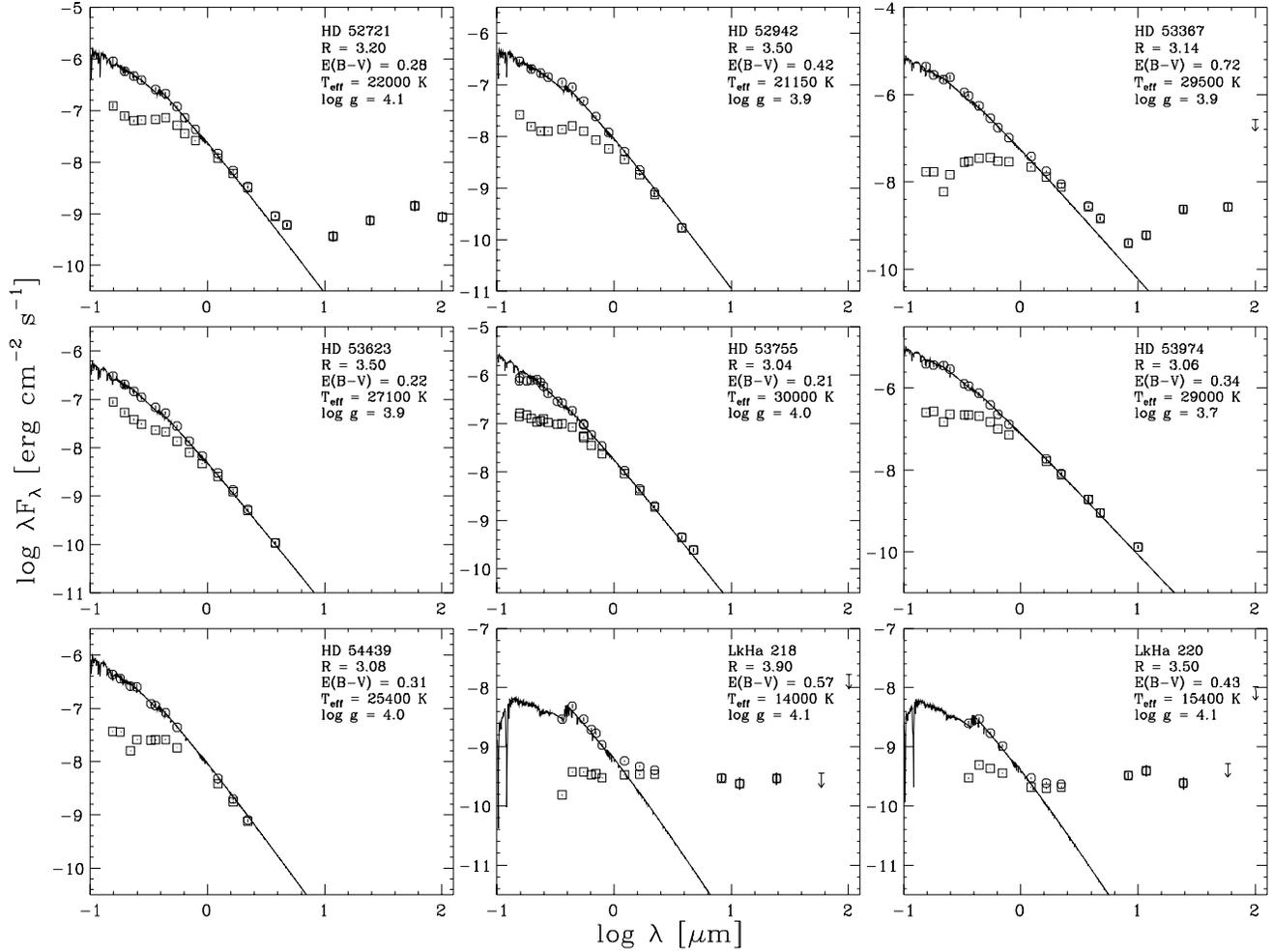,height=13.0cm,angle=0}}
\caption[]{Observed (squares) and extinction-corrected (circles) spectral 
 energy distributions of
 the nine programme stars.  The solid lines show Kurucz (1991) models for 
 the stellar photosphere, fitted to the extinction-corrected SED.}
\end{figure*}

\section{Observations}
   High resolution profiles of H$\alpha$, Na\,{\sc i}\,D, He\,{\sc i} 
(5876~\AA), Ca\,{\sc ii}\,K and two lines of the 
Ca\,{\sc ii}\,(2) red triplet for HD~52721 and HD~53367 and
of H$\alpha$, Na\,{\sc i}\,D and He\,{\sc i} (5876~\AA) for HD~52942, HD~53623, 
HD~53755, HD~53974 and HD~54439 have been secured in 1994, 1995 and 1996 at the 
European Southern Observatory (ESO), La Silla, Chile with the 1.4~m Coud\'e 
Echelle Spectrograph (CES) on the Coud\'e Auxiliary Telescope (CAT).  Low 
resolution spectra
of HD~52721, HD~53367 and LkH$\alpha$~220 were  
taken in 1992 and 1993  with the Boller and Chivens spectrograph on the 1.5 m 
telescope at ESO.  All spectra were reduced with the usual steps of bias 
subtraction, flatfielding, background subtraction and spectral extraction, 
and wavelength and flux calibration.
Additional high resolution optical spectra of HD~52721 and HD~53367 were 
obtained in October 1989 and November 1991 at the Special Astrophysical 
Observatory (SAO) of the Russian Academy of Sciences
with the Main Stellar Spectrograph (MSS) on the 6-m reflector telescope. These 
spectra were processed on the automatic spidometer and microdensitometer at 
the Crimean Astrophysical Observatory. Details of all observations are 
given in Table~1.

 Photometric observations of the stars have been made at ESO in various 
pass-bands
($UBVRIJHKLM$ and Walraven $WULBV$; optical photometry not published before is 
given 
in Table~5) which, together with the {\it IRAS} point source fluxes (paper I) 
and UV
fluxes from the {\it ANS} and {\it TD1} catalogues (Wesselius et al. 1982; 
Thompson et al. 1978), permitted us to 
determine the spectral energy distribution (SED) of each star.
For HD~52721 (GU CMa), HD~53367 (V570 Mon), LkH$\alpha$~218 (HT CMa) and 
LkH$\alpha$~220 (HU CMa)
we also constructed the light curves and colour-magnitude diagrams from the 
photometric
sequences obtained by the observers at Van Vleck Observatory (Herbst et al. 
1994), at Mt. Maidanak Observatory (Kilyakov \& Shevchenko 1976) 
and at Corralitos Observatory (Halbedel 1989, 1991, 1999).
High and low resolution ultraviolet spectra were taken from the Final 
Reduction Archive of the {\it IUE} (processed by NEWSIPS) at the ESA 
Satellite ground station in Villafranca del Castillo (Madrid). 
\begin{table}
\centering
\caption{Properties of programme stars.}
\tabcolsep0.08cm
\small
\begin{tabular}{@{}lllcccc}
\hline
Star       & Var. Name  & Sp. T.     &$d_{\rm ph}$& $V_J$  & $v \sin i$ & 
$v_{\rm rad}$\\
           &            &            & [pc]     & [mag]  & [km~s$^{-1}$] & 
[km~s$^{-1}$]\\
\hline 
HD 52721   &   GU CMa   & B1\,Ve      & ~600     & 6.40--6.80 &  350--450 & 
+9--+22\\
           &            & B1\,IVe     & ~800	 &            &	     & \\
HD 53367   &   V750 Mon & B0\,IVe     & ~775     & 6.90--7.20 &    25--50 & 
+11--+24\\
           &		& B0\,IIIe    & 1050	 &            &	     & \\
LkH$\alpha$ 218 & HT CMa & B6\,Ve     & ~900     &11.61--12.05&     50    & \\
LkH$\alpha$ 220 & HU CMa & B5\,Ve     & 1090     &11.60--12.19&     75    & \\
HD 52942   &   FZ CMa   & B2\,IV      & ~950     & 8.05--8.44 &    370    & \\
HD 53623   &		& B1\,V       & 1150     & 7.93--8.01 &	     & +30\\
HD 53755   &   V569 Mon & B0.5\,IV    & 1010     & 6.40--6.53 &    410    & 
+16\\
HD 53974   &   FN CMa   & B0\,II      & 1025     & 5.38--5.43 &    150    & 
+31\\
HD 54439   &		& B1\,V       & ~960     & 7.68       &    260    & \\
\hline
\end{tabular}
\end{table}

\section{Spectroscopic Results}
\subsection{The Photospheres}
 Spectroscopic classifications from the visual spectral range of the nine stars have been published 
frequently and are collected in Table 2. For the early B-type stars the photospheric information
from the visual spectrum is limited to some lines of Mg\,{\sc ii} and He\,{\sc i}, which may have resulted in 
some uncertainties in the spectral classifications. In the UV, however, more purely photospheric 
lines can be found: He\,{\sc ii} (1640~\AA), C\,{\sc iii} (1247~\AA), N\,{\sc iii} (1747, 1751~\AA), 
Si\,{\sc iii} (1299, 1417~\AA) and many lines of Fe\,{\sc iii}, Fe\,{\sc iv} and Fe\,{\sc v} 
(see e.g. Prinja 1990). In the {\it IUE} archive high resolution spectra
of four members of CMa R1 are available: HD~52721, HD~53367, HD~53755 and HD~53974.
We have compared these spectra with those in the UV spectral atlasses for B star classification 
by Rountree \& Sonneborn (1993) and Snow et al. (1994). 

From this comparison we derive the 
types B1\,IV--V for HD~52721 (comp. stars: HD~37303 B1\,V with  $v \sin i$ = 260 km~s$^{-1}$ and 
HD~108483 B2\,V with $v \sin i$ = 245 km~s$^{-1}$), B0\,IV,V for HD~53367 (comp. stars: HD~36512 
B0\,V with $v \sin i$ = 20 km~s$^{-1}$  and HD~34816 B0.5\,IV with $v \sin i$ = 67 km~s$^{-1}$) 
and B0.5\,III,IV for HD~53755 and HD~53974 (comp. stars: HD~219188 B0.5\,III with 
$v \sin i$ = 265 km~s$^{-1}$ and $\delta$~Sco (HD~143275) B0.3\,IV with $v \sin i$ = 180 km~s$^{-1}$).  
In this procedure approximate values of $v \sin i$ =  400, 30, 410 and 150 km~s$^{-1}$  have been 
taken into account for the stars HD~52721, HD~53367, HD~53755 and HD~53974 respectively 
(Uesugi \& Fukuda 1976; Finkenzeller 1985; B\"ohm \& Catala 1995). From the high 
ionization lines in the comparison spectra 
it also becomes clear that HD~53755 and HD~53974 have fast outflows. 
Our classifications from the UV spectra are close to those of the spectral 
and luminosity types from the visual spectra which were obtained earlier. 
Note that when we use the $M_V$ values of Schmidt-Kaler (1982), these classifications 
are consistent with a distance of $\sim$ 1 kpc (close to the value of 1050~pc, derived for CMa R1 
in paper I).
\begin{figure}
\centerline{\psfig{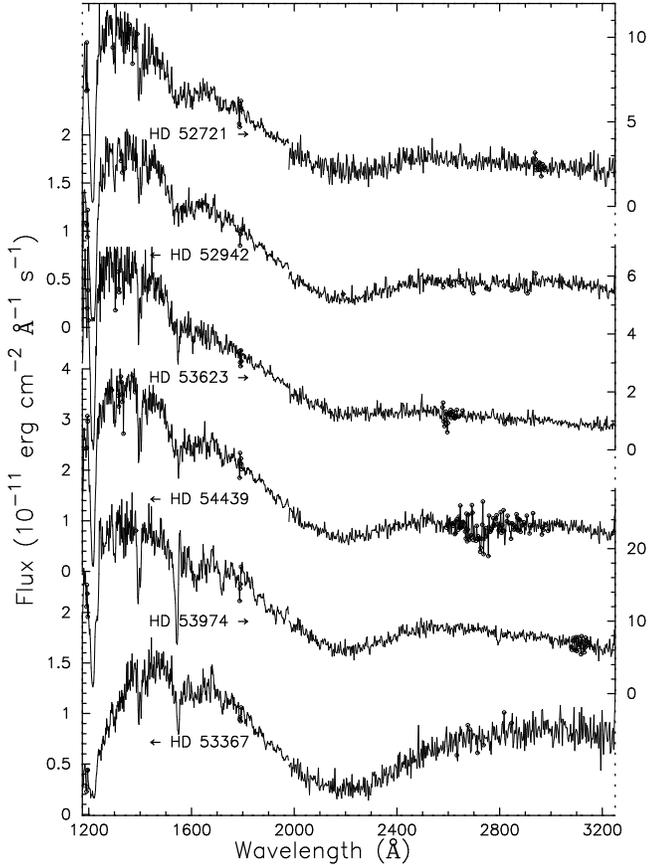}}
\caption[]{Low resolution {\it IUE} spectra of 6 stars. The arrows near the
stellar names point to the corresponding flux scales.}
\end{figure}
\begin{figure}
\centerline{\psfig{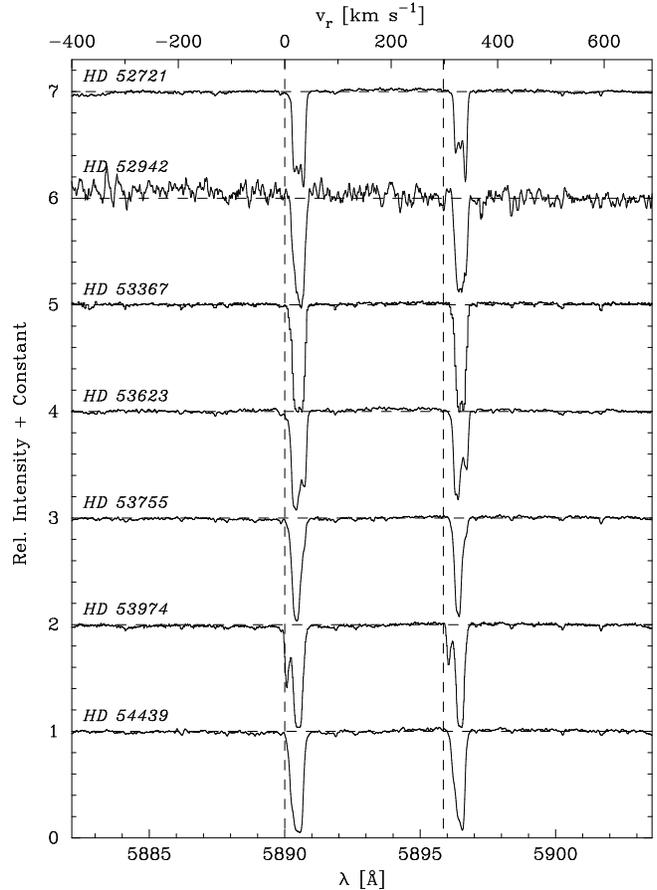}}
\caption[]{Na\,{\sc i}\,D profiles of 7 stars.  The horizontal and vertical 
dashed lines
 show the estimated continuum level and the Na\,{\sc i}\,D$_{2,1}$ rest 
wavelengths, respectively.}
\end{figure}

\subsection{The Outer Circumstellar Regions}
   Let us now inspect the information on the circumstellar dust and gas of the 
nine stars.
Fig.~1 shows the spectral energy distributions (SEDs) of the stars. These 
distributions are
fitted by the calculated Kurucz (1991) model distributions for the spectral and 
luminosity
class of  each star, together with the observed $E(B-V)$. In order to get a 
good agreement
in the visual and UV interval, one has to vary the extinction law, 
characterised by the
value of $R_V$ (the ratio of total absorption in the photometric $V$-band over 
the colour
excess $E(B-V)$).  The best fitting values of $R_V$, as well as the stellar 
parameters used for these fits, are included in Fig.~1.

The colour excess of HD~53367 is exceptionally large compared to those of 
the other stars in CMa R1. This is also expressed by the 2200~\AA\ absorption 
in the low
resolution UV spectrum of this star, which is much stronger then those in the 
corresponding
spectra of the other stars (Fig.~2). From Fig.~1 it is clear that only for the 
four H$\alpha$
emission line stars HD~52721, HD~53367, LkH$\alpha$~218 and LkH$\alpha$~220 the 
extinction
corrected SED has an excess flux at wavelengths longward 1.58~$\mu$m. Longward 
about 4~$\mu$m,
the excess for LkH$\alpha$~218 and LkH$\alpha$~220 increases strongly and can 
only be explained
by thermal re-emission from nearby hot circumstellar dust. 
For HD~52721 and HD~53367 the infrared excesses longward 4~$\mu$m are smaller 
and could be due to the presence of cooler circumstellar dust at larger 
distances from these stars, or to the contribution of cool stellar companions 
(see Sect. 4.2). The excesses at shorter wavelengths are much smaller and could 
be contributed by free-free emission in the inner circumstellar regions 
(see Sect. 3.3).

   In order to find more information about the gas component of the 
circumstellar medium of
the stars we will first discuss the high resolution profiles of the Na\,{\sc 
i}\,D (Fig.~3)
and Ca\,{\sc ii}\,K lines (Fig.~4c). For HD~52721 we find equal displacements 
(+19 km~s$^{-1}$)
and FWHM's (22 km~s$^{-1}$) for the Ca\,{\sc ii}\,K and Na\,{\sc i}\,D lines.  
For HD~53367 we find similar values (displaced +17 km~s$^{-1}$ and 24 
km~s$^{-1}$) for
Ca\,{\sc ii}\,K 
and Na\,{\sc i}\,D. This indicates that the resonance line of  Ca\,{\sc ii} is 
formed
in the same region as those of Na\,{\sc i}.  The shape of the Na\,{\sc i}\,D 
profiles (Fig.~3)
shows the 
presence of more than one component, which could be contributed by interstellar clouds 
in the CMa
association or else from circumstellar gas. The redshifts of HD~52721 and 
HD~53367 are within
the range of their radial velocity variations (Abt \& Biggs 1972).  The main 
components
are redshifted by about 19 km~s$^{-1}$ for HD~53974, HD~53755 and HD~54439, 
which is within the
range
of the radial velocities measured for the double stars HD~53974 and HD~53755 
(Abt \& Biggs 1972).
For HD~53623 we measure +12 km~s$^{-1}$ for the main (blue) components and +27 
km~s$^{-1}$ for the second
component. The radial velocity from the literature (Abt \& Biggs 1972) is +30 
km~s$^{-1}$. There
are no indications that this star is a binary. HD~52942 is a double line 
eclipsing binary with equal 
mass companions (Moffat \& Vogt 1974), but we see only one Na\,{\sc i}\,D 
doublet in our (noisy) 
spectrum. The difference in redshift may be hidden by the noise, but the 
average value
(+15 km~s$^{-1}$) can only be understood if the absorption is formed at some 
distance of the
binary system.  From the general agreement between the Na\,{\sc i}\,D redshifts 
and the stellar
radial velocities we conclude that the Na\,{\sc i}\,D doublet is formed in the 
(wide) circumstellar region of the stars.
From the EWs of the Na\,{\sc i}\,D doublet lines we can evaluate the 
column density of the foreground gas with  the Str\"omgren (1948) doublet-ratio 
method.
In Table 3 we give the EWs of the Na\,{\sc i}\,D lines and the derived column 
densities
of Na\,{\sc i}\,D.  From $N$(Na\,{\sc i}) and by assuming a galactic abundance 
ratio
$N$(H)/$N$(Na\,{\sc i}) of $5 \times 10^5$ we can derive colour excesses for an 
average
interstellar dust/gas ratio with the empirical relation of Hobbs (1974): 
$N$(Na\,{\sc i}) = $1.7 \times 10^{14}$ $\times$ $E(B-V)^{1.8}$.  The results 
are listed in column 5 of Table 3 and can be compared with the values 
of $E(B-V)$ (column 6), derived from the photometric observations by fitting 
the SEDs. From this comparison it is
clear that the interstellar dust gives significantly lower colour excesses than 
those obtained
directly from the photometry (see also Paper I Sect. 3.1). This can be 
understood if the
dust/gas ratio in the circumstellar and intra-association cloud regions is 
higher
than in the general interstellar foreground.  From Table 3 we see that this 
ratio will be especially high for HD~53367. 
\begin{table*}
\centering
\caption{Comparison of interstellar colour excess, derived from 
 the Na\,{\sc i}\,(1) doublet with observed total colour excess.}
\small
\begin{tabular}{@{}lccccc}
\hline
Star	     &	EW (Na\,D$_1$)& EW2/EW1 & $\log N(\mbox{Na})$ & $E(B-V)_{\rm Na}$ & 
$E(B-V)_{\rm obs.}$\\
\hline
HD 52721     &	 0.308	&  1.32	  &  12.72  &   0.145  &   0.28\\
HD 52942     &	 0.393	&  1.18	  &  13.05  &  	0.222  &   0.42\\
HD 53367     &	 0.461	&  1.17	  &  13.18  &  	0.259  &   0.72\\
HD 53623     &	 0.375	&  1.30	  &  12.84  &  	0.170  &   0.22\\
HD 53755     &	 0.311	&  1.24	  &  12.85  &  	0.171  &   0.21\\
HD 53974     &	 0.410	&  1.22	  &  13.07  &  	0.226  &   0.34\\
HD 54439     &	 0.399	&  1.21	  &  13.00  &  	0.207  &   0.31\\
LkH$\alpha$ 218$^\dagger$ &	 0.117	&  1.23	  &  12.42  &  	0.098  &   0.57\\
LkH$\alpha$ 220$^\dagger$ &	 0.141	&  1.03	  &  13.71  &  	0.513  &   0.43\\
\hline
\end{tabular}
\noindent
\flushleft
$^\dagger$The EWs for LkH$\alpha$ 220 and LkH$\alpha$ 218
are less accurate than those for the other stars, because they were measured
from the figures in the paper of Finkenzeller \& Mundt (1984).
\end{table*}

     We have inspected the high resolution {\it IUE} spectra of HD~52721, 
HD~53367, HD~53755 and HD~53974. In these 
spectra one finds a large number of narrow lines with zero excitation energy of 
their lower levels, which will 
be formed in the interstellar foreground: e.g. at 2851~\AA\ (Mg\,{\sc i}\,(1)), 
2026~\AA\ (Mg\,{\sc i}\,(2)), 2795~\AA, 2802~\AA\ ((Mg\,{\sc ii}\,(1)), 
2593~\AA\ (Mn\,{\sc ii}\,(1)), 2585~\AA\ (Fe\,{\sc ii}\,(1)), 2382~\AA\ 
(Fe\,{\sc ii}\,(2)), 1526~\AA\ (Si\,{\sc ii}\,(2)). 
In contrast, the number of low excitation lines 
in these spectra is small and these lines are very weak. We note that only 
a few lines  (2550.8~\AA\ of Fe\,{\sc ii}\,(240), 2556.6~\AA\ of Mn\,(20), 
2666.6~\AA\ of Fe\,{\sc ii}\,(263) and 2724.5~\AA\ of Mn\,{\sc ii}\,(33)) with
lower levels up to $\sim$ 3.5~eV are present in the spectra of HD~52721 and 
HD~53974 but
are missing in those of HD~53367 and HD~53755. Since the latter two stars 
are not far apart from each other in CMa R1, their common lack of these 
low excitation lines could
be due to the conditions in their common foreground interstellar clouds. 
The scarceness and weakness of the low excitation lines indicate that the 
four stars which we inspected in the UV have only a very small amount of 
circumstellar gas, which is consistent with the high dust/gas ratio in 
the circumstellar region.  Only one of the stars, HD~53367,
shows the possible presence of the Mg\,{\sc ii}\,(3) doublet (2790.77 and 
2797.99~\AA) in one of its spectra.
\begin{figure*}
\centerline{\psfig{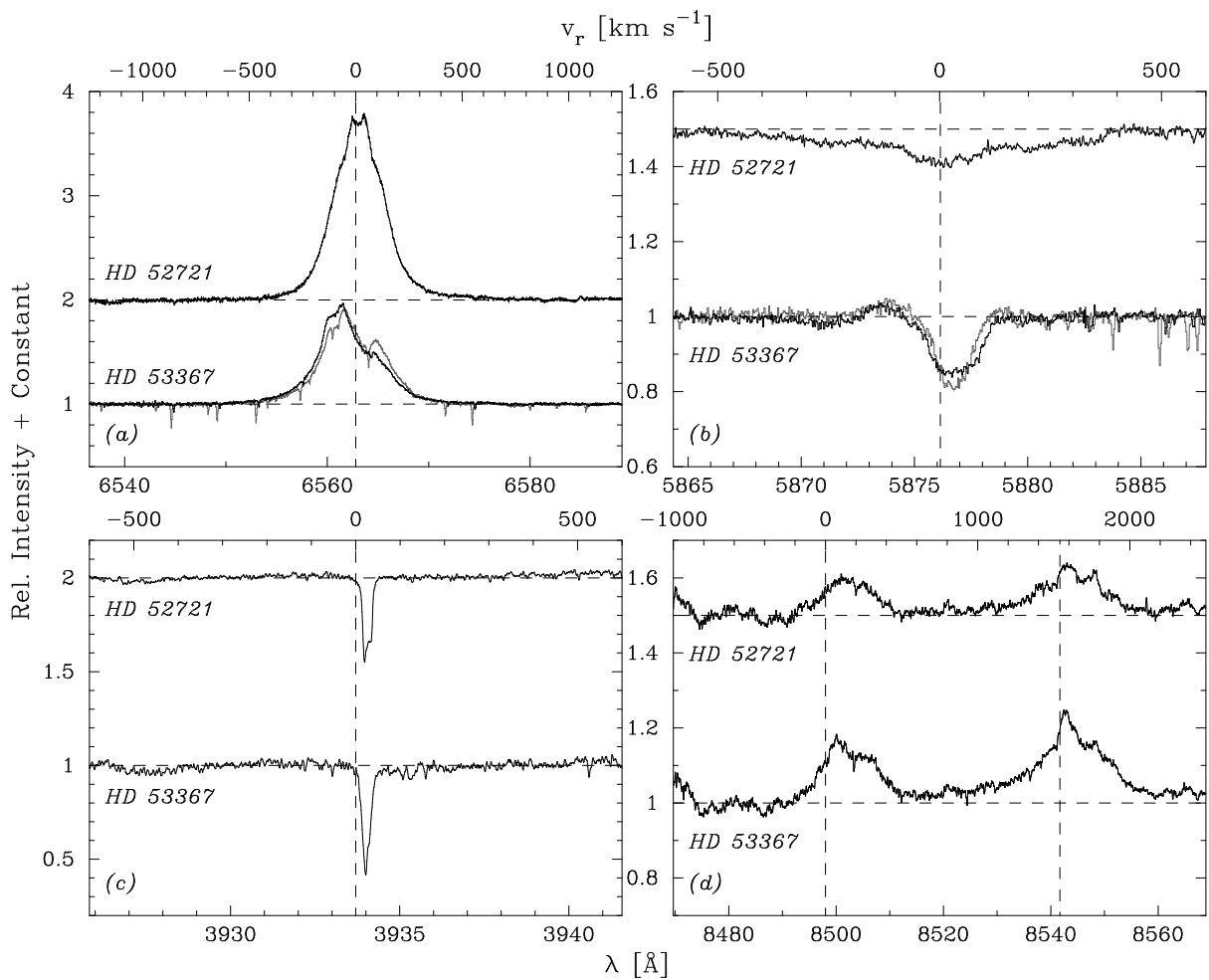}}
\caption[]{High resolution profiles for HD~52721 and HD~53367. (a) H$\alpha$, 
(b)
 He\,{\sc i} (5876~\AA), (c) Ca\,{\sc ii}\,K and (d) Ca\,{\sc ii}\,(2). For 
 HD~53367 the black and grey profiles of H$\alpha$ and He\,{\sc i} correspond 
 to the two exposures, one month apart (see Table~1).}
\end{figure*}
\begin{figure*}
\centerline{\psfig{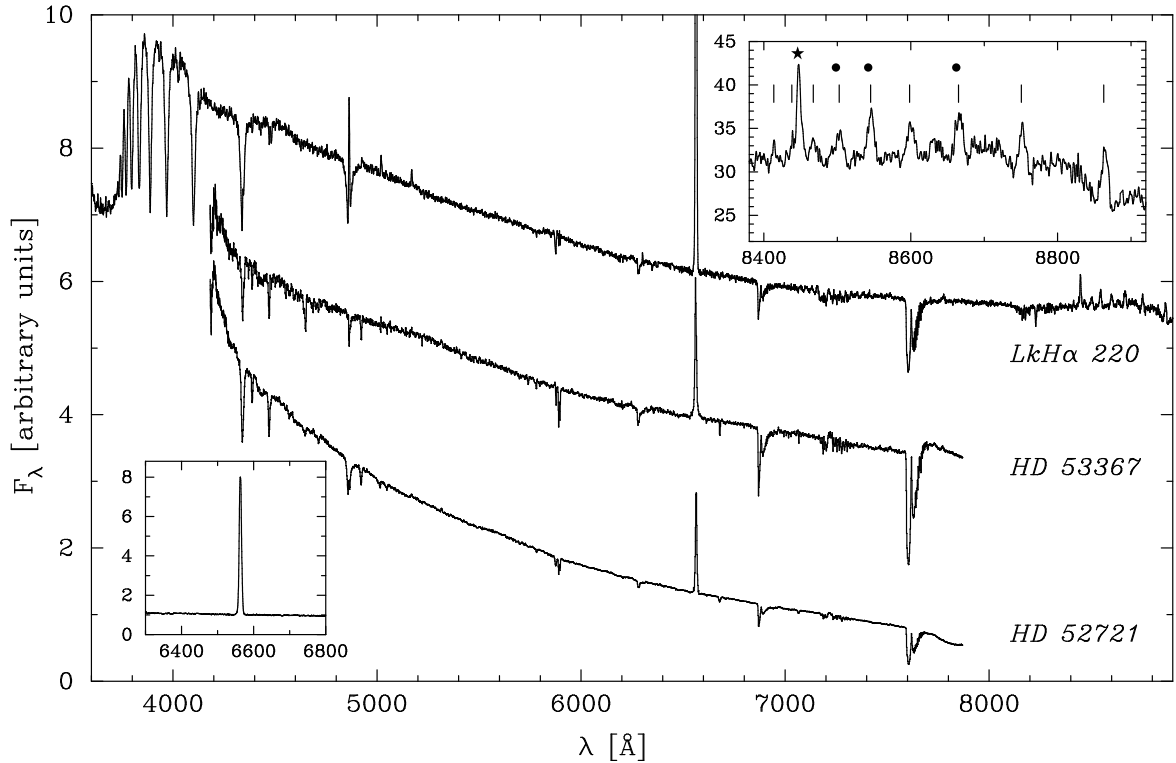}}
\caption[]{Low resolution spectra of LkH$\alpha$~220, HD~52721 and HD~53367. 
Insets
 show H$\alpha$ and the red spectrum of LkH$\alpha$~220 in more detail. Lines, 
dots
 and stars in the inset mark the positions of lines of H\,{\sc i}, Ca\,{\sc ii} 
and
 O\,{\sc i}, respectively (see Table 4).}
\end{figure*}
\begin{table*}
\centering
\caption{Equivalent Widths (EWs) and Full Width Half Max. (FWHMs) of four 
emission-line stars in
CMa R1.}
\small
\begin{tabular}{@{}lllcccrl}
\hline
Star     & \multicolumn{1}{c}{$\lambda$ [\AA]}& Line       & EW [\AA] & Error 
[\AA]& FWHM [km~s$^{-1}$] & \multicolumn{1}{c}{Date} & Ref.\\
\hline   
HD 52721 & 3933.6  & Ca\,{\sc ii}\,K   &    +0.14   &   0.03   &    25    &   
14 Dec. 1996 & This paper\\
    	 & 5875.6  & He\,{\sc i}       &    +0.67   &   0.10   &   490    &   13 
Dec. 1996 & This paper\\
    	 & 6562.8  & H$\alpha$         &  $-$5.2    &          &   330    &   14 
Nov. 1981 & Finkenzeller \& Mundt (1984)\\
    	 & 6562.8  & H$\alpha$         &  $-$9.4    &          &          &   31 
Jan. 1983 & Andrillat (1983)\\
    	 & 6562.8  & H$\alpha$         &  $-$10--14 &          &   300    & 4--9 
Mar. 1985 & Praderie et al. (1991)\\
    	 & 6562.8  & H$\alpha$         &  $-$8.8    &          &          &   23 
Dec. 1991 & Corcoran \& Ray (1998)\\
    	 & 6562.8  & H$\alpha$         &  $-$14     &          &          &   11 
Jan. 1995 & Oudmaijer \& Drew (1999)\\
    	 & 6562.8  & H$\alpha$         &  $-$9.9    &   0.15   &          &   11 
Dec. 1996 & This paper\\
    	 & 7773    & O\,{\sc i}        &  $-$0.24   &          &   311    &      
Dec. 1987 & Hamann \& Persson (1992)\\
    	 & 8446.5  & O\,{\sc i}        &  $-$2.06   &          &          &      
Dec. 1987 & Hamann \& Persson (1992)\\
    	 & 8465    & P17               &  $-$0.35   &          &   335    &   22 
Jan. 1983 & Andrillat et al. (1988)\\
    	 & 8498.02 & Ca\,{\sc ii}\,(2) &  $-$0.46   &   0.07   &   225    &   15 
Dec. 1996 & This paper\\
    	 & 8502.49 & P16               &  $-$0.34   &   0.06   &   225    &   15 
Dec. 1996 & This paper\\
    	 & 8500    & Ca\,{\sc ii} + P16&  $-$0.72   &          &          &   22 
Jan. 1983 & Andrillat et al. (1988)\\
    	 & 8542.09 & Ca\,{\sc ii}\,(2) &  $-$0.75   &   0.10   &   225    &   15 
Dec. 1996 & This paper\\
    	 & 8548.39 & P15               &  $-$0.37   &   0.07   &   225    &   15 
Dec. 1996 & This paper\\
    	 & 8545    & Ca\,{\sc ii} + P15&  $-$0.55   &          &          &   22 
Jan. 1983 & Andrillat et al. (1988)\\
    	 & 8598    & P14               &  $-$0.35   &          &          &   22 
Jan. 1983 & Andrillat et al. (1988) \\
\\
HD 53367 & 3933.6  & Ca\,{\sc ii}\,K   &    +0.19   &   0.06   &    25    &   
14 Dec. 1996 & This paper\\
 	 & 5875.6  & He\,{\sc i}       &    +0.34   &   0.03   &   121    &   16 Dec. 
1994 & This paper\\
 	 & 5875.6  & He\,{\sc i}       &  $-$0.03   &   0.01   &    70    &   16 Dec. 
1994 & This paper\\
 	 & 5875.6  & He\,{\sc i}       &    +0.37   &   0.04   &    91    &   16 Jan. 
1995 & This paper\\
 	 & 5875.6  & He\,{\sc i}       &  $-$0.07   &   0.03   &    66    &   16 Jan. 
1995 & This paper\\
 	 & 5875.7  & He\,{\sc i}       &    +0.42   &          &    70:   &   14 Nov. 
1981 & Finkenzeller \& Mundt (1984)\\
 	 & 6562.8  & H$\alpha$         &  $-$19.4   &          &   300    &           
1974 & Garrison \& Anderson (1977)\\
 	 & 6562.8  & H$\alpha$         &  $-$14     &          &          &   22 Oct. 
1981 & Andrillat (1983)\\
 	 & 6562.8  & H$\alpha$         &  $-$11     &          &   210    &   14 Nov. 
1981 & Finkenzeller \& Mundt (1984)\\
 	 & 6562.8  & H$\alpha$         &  $-$14.3   &          &   340    &   20 Nov. 
1987 & Halbedel (1989)\\
 	 & 6562.8  & H$\alpha$         &  $-$16.2   &          &   325    &      Dec. 
1987 & Hamann \& Persson (1992) \\
 	 & 6562.8  & H$\alpha$         &  $-$5.0    &          &          &    1 Jan. 
1992 & Corcoran \& Ray (1998)\\
 	 & 6562.8  & H$\alpha$         &  $-$6.1    &   0.12   &   250    &   16 Dec. 
1994 & This paper\\
 	 & 6562.8  & H$\alpha$         &  $-$6.1    &   0.12   &   250    &   15 Jan. 
1995 & This paper\\
 	 & 6562.8  & H$\alpha$         &  $-$14     &          &          &    1 Jan. 
1997 & Oudmaijer \& Drew (1999)\\
 	 & 7773    & O\,{\sc i}        &  $-$0.37   &          &   424    &      Dec. 
1987 & Hamann \& Persson (1992)\\
 	 & 8446.5  & O\,{\sc i}        &  $-$2.68   &          &          &      Dec. 
1987 & Hamann \& Persson (1992)\\
 	 & 8465    & P17               &  $-$1.11   &          &          &   22 Jan. 
1983 & Andrillat et al. (1988)\\
 	 & 8498.02 & Ca\,{\sc ii}\,(2) &  $-$0.83   &   0.08   &   210    &   15 Dec. 
1996 & This paper\\
 	 & 8502.49 & P16               &  $-$0.53   &   0.05   &   210    &   15 Dec. 
1996 & This paper\\
 	 & 8500    & Ca\,{\sc ii} + P16&  $-$1.58   &          &          &   22 Jan. 
1983 & Andrillat et al. (1988)\\
 	 & 8542.09 & Ca\,{\sc ii}\,(2) &  $-$1.16   &   0.12   &   210    &   15 Dec. 
1996 & This paper\\
 	 & 8548.39 & P15               &  $-$0.74   &   0.07   &   210    &   15 Dec. 
1996 & This paper\\
 	 & 8545    & Ca\,{\sc ii} + P15&  $-$1.34   &          &          &   22 Jan. 
1983 & Andrillat et al. (1988)\\
 	 & 8598    & P14               &  $-$1.54   &          &          &   22 Jan. 
1983 & Andrillat et al. (1988) \\
\\
LkH$\alpha$ 218 & 6562.8  & H$\alpha$         &  $-$24.4   &          &          &   
14 Jan. 1977 & Cohen \& Kuhi  (1979)\\
  	 & 6562.8  & H$\alpha$         &  $-$19     &   0.5    &   300    &   14 
Nov. 1981 & Finkenzeller \& Mundt (1984)\\
  	 & 6562.8  & H$\alpha$         &  $-$26     &   1.3    &          &   30 
Jan. 1986 & B\"ohm \& Catala (1995)\\
  	 & 6562.8  & H$\alpha$         &  $-$32     &          &          &   25 
Dec. 1991 & Corcoran \& Ray (1998)\\
  	 & 6562.8  & H$\alpha$         &  $-$20     &          &          &   31 
Dec. 1996 & Oudmaijer \& Drew (1999)\\
  	 & 8545    & Ca\,{\sc ii} + P15&  $-$2.1    &   0.4    &          &   30 
Jan. 1986 & B\"ohm \& Catala (1993)\\
\\
LkH$\alpha$ 220 & 6562.8  & H$\alpha$         &  $-$52.5   & 	       &          &   14 
Jan. 1977 & Cohen \& Kuhi (1979)\\
         & 6562.8  & H$\alpha$         &  $-$35     & 	0.5    &   280    &   14 
Nov. 1981 & Finkenzeller \& Mundt (1984)\\
         & 6562.8  & H$\alpha$         &  $-$67     & 	3.3    &          &   30 
Jan. 1986 & B\"ohm \& Catala (1995)\\
         & 6562.8  & H$\alpha$         &  $-$74     & 	0.7    &   300    &    1 
Jan. 1994 & This paper\\
         & 8446.5  & O\,{\sc i}        &  $-$3.3    & 	0.3:   &   250:   &    1 
Jan. 1994 & This paper\\
         & 8465    & P17               &  $-$0.70   & 	0.3:   &   330:   &    1 
Jan. 1994 & This paper\\
         & 8500    & Ca\,{\sc ii} + P16&  $-$0.98   & 	0.3:   &   350:   &    1 
Jan. 1994 & This paper\\
         & 8545    & Ca\,{\sc ii} + P15&  $-$1.80   & 	0.3:   &   350:   &    1 
Jan. 1994 & This paper\\
         & 8598    & P14               &  $-$1.28   & 	0.3:   &   350:   &    1 
Jan. 1994 & This paper\\
         & 8662    & Ca\,{\sc ii} + P13&  $-$1.75   & 	0.3:   &   400:   &    1 Jan. 1994 & This paper\\
         & 8750    & P12               &  $-$1.50   &   0.3:   &   420:   &    1 Jan. 1994 & This paper\\
         & 8862    & P11               &  $-$2.00   &   0.3:   &   430:   &    1 Jan. 1994 & This paper\\
\hline
\end{tabular}
\end{table*}

\subsection{The Inner Circumstellar Regions}
\subsubsection{The B-emission line stars: HD~52721, HD~53367, LkH$\alpha$~218 
and LkH$\alpha$~220}
    Of special interest are the emission lines of H$\alpha$ (Fig.~4a), the red 
Ca\,{\sc ii}\,(2)
lines (Fig.~4d) and the emission component in the He\,{\sc i} lines in the 
visual range of the
spectra of HD~52721 and HD~53367 (Fig.~4b). 
  In high resolution, H$\alpha$  shows emission profiles (Fig.~4a) which are 
symmetrical
for HD~52721 and asymmetrical (distorted by a red-shifted absorption component) 
for HD~53367.
For the first star the shape does not vary much on short (days) and long (yrs) 
timescales
(see Sect. 4). For HD~53367 small shape variations in the red wing occur on 
time-scales
of one month or less (Fig.~4a) and large shape variations over the whole 
profile occur
over time-scales of years. Because of its very low $v \sin i$ ($\approx$ 30 
km~s$^{-1}$)
we could correct 
the H$\alpha$ profile of HD~53367 for its underlying photospheric profile 
(assumed to
be that of a Kurucz model H$\alpha$ profile for $T_{\rm eff}$ = 30,000~K and 
$\log g$ = 3.5)
and conclude that the outflow velocity is 100--150 km~s$^{-1}$.

For HD~52721 with its 
large $v \sin i$ (350--400 km~s$^{-1}$) we have no comparison profile, but the 
observed
profile indicates that the outflow velocity of this star is less than 100 
km~s$^{-1}$. The
parameters of H$\alpha$ of the four stars are given in Table 4, which shows 
that they are variable
in time.  For HD~53367 we measured an EW of $-$6.1~\AA, which strongly differs 
from $-$14.0~\AA\
(Andrillat 1983) and $-$16.2~\AA\ (Hamann \& Persson 1992), but is close to the 
$-$5.0~\AA, measured by Corcoran \& Ray (1998) for their profile of Jan. 1, 
1992. We measure a
FWHM of 260 km~s$^{-1}$ for HD~52721 and 280 km~s$^{-1}$ (or 
310 km~s$^{-1}$ if we account for the presence of an absorption component) for 
HD~53367. For
LkH$\alpha$~218 and LkH$\alpha$~220, emission profiles of H$\alpha$ on Nov. 15, 
1981 have been
presented by Finkenzeller \& Mundt (1984). These profiles have double peaked 
shapes with
blue-shifted ($\approx$ $-$70 km~s$^{-1}$) absorptions.  The variability of 
H$\alpha$
is further discussed in Sect. 4.  
\begin{figure}
\centerline{\psfig{figure=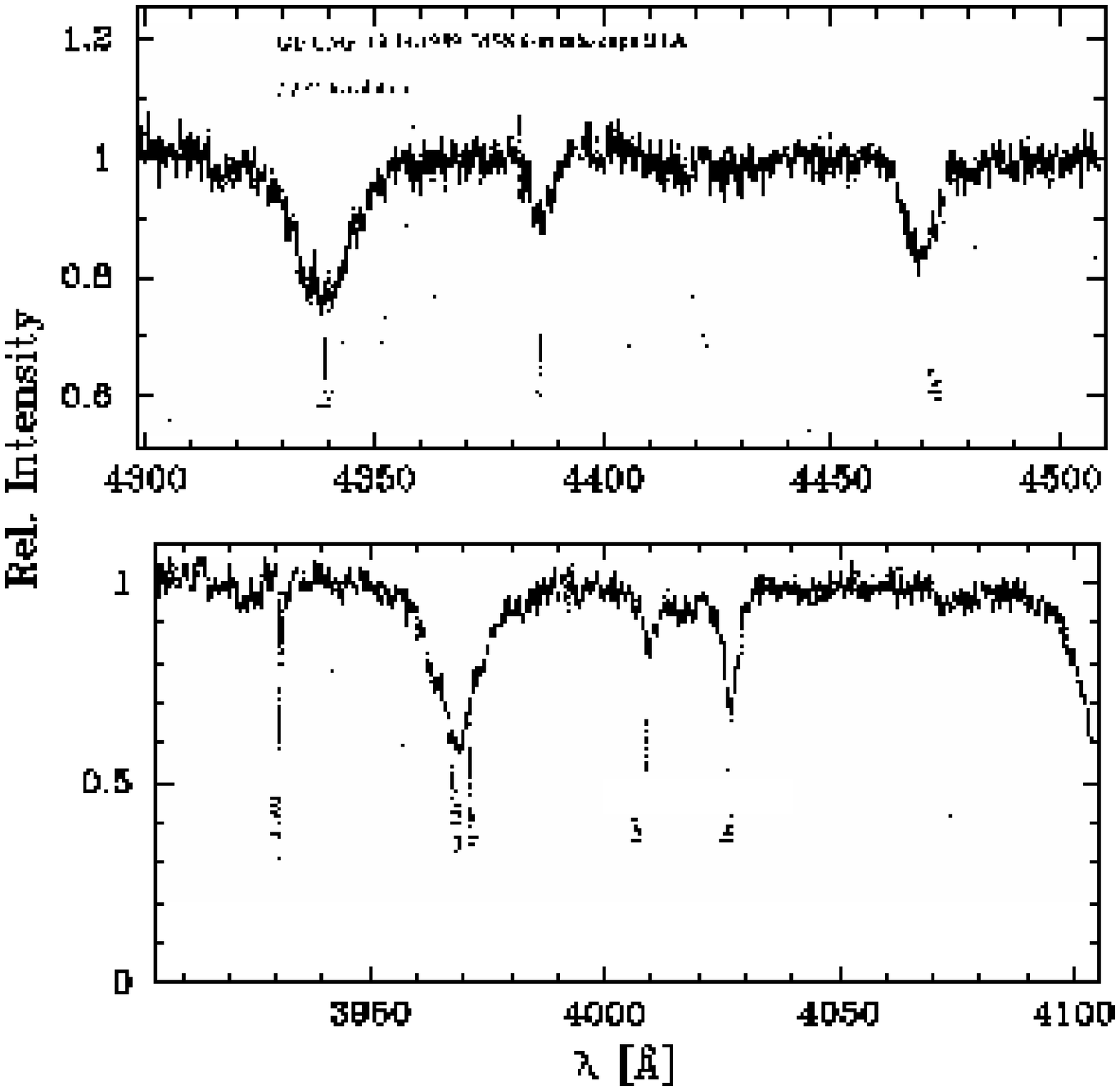,width=8.5cm,angle=0}}
\caption[]{High resolution spectral intervals of HD~52721 taken on Oct. 14, 1989 
 with the Main Stellar Spectrograph on the 6-m BTA telescope of the SAO. 
 The upper panel shows profiles of H$\gamma$ and He\,{\sc i} (4387.9 and 
 4471.5~\AA). The lower panel depicts those of 
 H$\varepsilon$, Ca\,{\sc ii}\,H and K and He\,{\sc i} (4009.3 and 4026.2~\AA). 
 }
\end{figure}

For HD~52721 and HD~53367 we have derived volume emission measures from the 
fill-in
of the Balmer discontinuity by bound-free and free-free emission in the 
circumstellar envelope.
This fill-in could be determined by comparing the Balmer jump, derived from 
photometric
observations 
on the Walraven system (which has two pass bands shortward and two passbands 
longward
the Balmer limit at 3704~\AA), with the model-predicted Balmer discontinuity 
(e.g. from the non-LTE
models of Mihalas). The observed Walraven magnitudes and the derived Balmer 
fill-in
$\Delta D_b$ are given in Table~5. With the method of Garrison (1978) and the 
assumption that
$T_{\rm cs}$ $\approx$ 10,000~K, we then derive volume emission measures 
$E$ = $\int n_e n_i \mbox{d}V/R_\star^2$ of 
$6.1 \times 10^{36}$~cm$^{-5}$ for HD~52721 and $1.3 \times 10^{37}$~cm$^{-5}$ 
for HD~53367.
If we assume an electron density of $10^{12}$~cm$^{-3}$ in the circumstellar 
envelope region close
to the stars we estimate with the formula of Garrison (1978) that the mass 
outflow in the
circumstellar envelope close to the star is $9.2 \times 
10^{-7}$~M$_\odot$~yr$^{-1}$ for HD~52721 and 
$4.2 \times 10^{-6}$~M$_\odot$~yr$^{-1}$ for HD~53367.  
\begin{table}
\centering
\caption{Walraven photometry of HD~52721 and HD~53367 and corresponding 
fill-in of Balmer decrement, emission measure and mass loss rate.}
\small
\begin{tabular}{@{}lcc}
\hline
                  &  HD~52721            &  HD~53367\\
\hline
Date               &  May 1990            &  25 Jan. 1981\\
$V_W$              &  0.153 $\pm$ 0.002   & $-$0.002 $\pm$ 0.002\\
$(V-B)_W$          &  0.036 $\pm$ 0.002   &  ~~0.202 $\pm$ 0.002\\
$(B-L)_W$          &  0.018 $\pm$ 0.002   &  ~~0.055 $\pm$ 0.002\\
$(B-U)_W$          &  0.016 $\pm$ 0.003   &  ~~0.043 $\pm$ 0.003\\
$(B-W)_W$          &  0.027 $\pm$ 0.003   &  ~~0.146 $\pm$ 0.003\\
$V_J$ [mag]        &  6.48                &  6.90\\
$\Delta D_b$ [mag] & 0.24                 &  0.30\\
EM [cm$^5$]        & $6.2 \times 10^{36}$ & $1.3 \times 10^{37}$\\
$v_{H\alpha}$ [km~s$^{-1}$]& $\approx$ 100 & $\approx$ 140\\
$\dot{M}$ [M$_\odot$~yr$^{-1}$]& $9.2 \times 10^{-7}$ & $4.2 \times 10^{-6}$\\
\hline
\end{tabular}
\end{table}

The high resolution profiles of the emission lines at 8500 and 8543~\AA\ 
(Fig.~4d) allow us to separate the 8498.02 and 8542.09~\AA\ lines of the 
Ca\,{\sc ii}\,(2) triplet from the partly blended Paschen lines 
P16 (8502.49~\AA) and P15 (8548.39~\AA). As a result of the blending,
the peaks are slightly shifted. For HD~52721 we have decomposed the 
profiles into two components
with a FWHM of 225 km~s$^{-1}$ each, which gives us EW(P16) = $-$0.34~\AA, 
EW(Ca\,{\sc ii} 8498) = $-$0.46~\AA, 
EW(P15)= $-$0.37~\AA\ and EW(Ca\,{\sc ii} 8542) = $-$0.75~\AA. Together we have 
$-$0.80~\AA\ for
the 8500 profile and $-$1.12~\AA\ for the 8543 profile, compared with 
$-$0.72~\AA\ and $-$0.55~\AA\
determined by Andrillat, Jaschek \& Jaschek (1988). Note that these authors 
give an
EW of $-$0.35~\AA\ to both unblended P14 (8598~\AA) and P17 (8465~\AA) lines. 
We therefore
may expect a similar EW for P15 and P16, which is very close to our results 
above.
The EWs of the blended  lines quoted by Andrillat et al. differ from our values 
in Table 4,
which may indicate a variability of the Ca\,{\sc ii}\,(2) fluxes. 

For HD~53367 we made similar 
decompositions with FWHM values of 205 km~s$^{-1}$. This gives EW(P16) = 
$-$0.56~\AA,
EW(Ca\,{\sc ii} 8498) = $-$0.83~\AA, EW(P15) = $-$0.74~\AA\ and EW(Ca\,{\sc ii} 
8542) = $-$1.16~\AA.
Andrillat et al. (1988) give EW(P14) = $-$1.54~\AA\ and EW(P17) = $-$1.11~\AA\ 
for this star.
This is consistent with our result above that the EW of H$\alpha$ in 1992 was 
very low
(much smaller than in Oct. 1981). The blended lines at 8500 and 8543~\AA\ were 
observed
by Andrillat et al. in Jan. 1983. Our most important conclusion from the high 
resolution
profiles of HD~52721 and HD~53367 is that the emission of the Ca\,{\sc ii}\,(2) 
lines is significant and has a FWHM of the order of 210 km~s$^{-1}$. 
Parameters of the Ca\,{\sc ii} lines are collected in Table 4.

Since Ca\,{\sc ii} cannot exist at the stellar photosphere the emission of 
Ca\,{\sc ii}\,(2) must originate 
from the envelope region. Hamann \& Persson (1992) have given arguments for the 
presence
of discs around Herbig Ae/Be stars with H$\alpha$ and Ca\,{\sc ii} emission. 
For early B-type stars these
discs are probably very similar to those of classical Be stars. The FWHM's of 
the Ca\,{\sc ii}\,(2)
lines suggest that Ca\,{\sc ii}\,(2) is emitted in an outer region of a 
circumstellar disc
(at a Kepler radius of 29~R$_\star$ for HD~53367 and 17.2~R$_\star$ for 
HD~52721), whereas the
H$\alpha$ emission could come from a region of the disc which is somewhat 
closer to the star
(at a Kepler radius of 13.5~R$_\star$ for HD~53367 and 12.4~R$_\star$ for 
HD~52721).
Here we have implicitly assumed that both star-discs are being observed more or 
less edge-on.  It is interesting to note that the low resolution red spectrum 
of LkH$\alpha$~220 (Fig.~5) also shows the presence of strong emission in 
the Ca\,{\sc ii}\,(2) triplet.

   The profiles of the O\,{\sc i} emission lines for HD~52721 and HD~53367 in 
the red part of the spectrum 
have been observed by Hamann \& Persson (1992).  The emission profiles of these 
lines are weak
and broad (Fig.~3 in their paper) and only the widths of the blended 8446~\AA\ 
O{\sc i}\,(4) line could 
be estimated (FWHM of  $\sim$ 300 km~s$^{-1}$). The authors report  EWs of 
$-$0.24~\AA\ and
$-$0.37~\AA\ for the unblended triplet O\,{\sc i}\,(1) line (7773~\AA) of 
HD~52721 and HD~53367 respectively 
and $-$2.06~\AA\ and  $-$2.68~\AA\ for the blended O\,{\sc i}\,(4) line 
(8446~\AA) of HD~52721 and HD~53367 respectively. 
The 8446~\AA\ O\,{\sc i}\,(4) emission is much stronger than that of the 
7773~\AA\ line,
which cannot completely be explained by the blend with P18 (8437~\AA). This 
O\,{\sc i}\,(4)
emission is common in classical Be stars and probably due to pumping 
of the $^3d$~$^3D_0$ state of
O\,{\sc i} by Ly$\beta$ (Bowen 1947). The fact that the 1302~\AA\ O\,{\sc 
i}\,(2) line
is in absorption, while for the same stars the O\,{\sc i}\,(4) transition, 
which feeds
the 1302~\AA\ transition, is in emission has led Oegerle, Peters \& Polidan 
(1983) to
argue that because the optical depth of line scattering in the resonance 
1302~\AA\ line
is much larger than in the subordinate 8446~\AA\ line, the 1304~\AA\ photons 
could be
scattered out the line-of-sight in equatorially viewed flattened Be envelopes.  
It would
suggest that the O\,{\sc i} emission is coming from flattened discs around 
HD~52721 and HD~53367 and 
that both discs are viewed edge-on. This suggestion seems to be confirmed by 
recent
H$\alpha$ polarization measurements (Oudmaijer \& Drew 1999) of HD~53367 which 
indicate that this
star is not observed pole-on. In the low resolution red spectrum of 
LkH$\alpha$~220
(Fig.~5) we can distinguish the emission in O\,{\sc i}\,(1) at 7773~\AA, and 
that in
O\,{\sc i}\,(4) at 8446.5~\AA, slightly blended with P18.

   The He\,{\sc i} (5876~\AA) absorption profile of HD~52721 (Fig. 4b) is very broad with a FWHM of 
$\approx$ 450 km~s$^{-1}$, approximately the same as for the He\,{\sc i} lines at 4026 and 4471~\AA\ 
in the spectrum of this star, obtained by Finkenzeller \& Mundt (1984).  This large width 
probably reflects the photospheric origin of these He\,{\sc i} lines in this fast rotating 
star. Inside the line profile of the 5876~\AA\ line a weak emission component 
can be seen at both sides of the line centre.The He\,{\sc i} 6678~\AA\ profiles 
of this star (Corporon \& Lagrange 1999) also have such an emission component, which shows 
variability in strength on a time-scale of less than one day. Such emission components have been 
observed more often in the He\,{\sc i} lines of B and Be stars and have been interpreted as CQE's
(central quasi-emission peaks), reduced absorption which can occur when the stellar disc is 
occulted by circumstellar gas moving in Kepler rotation around the star and when we observe 
the circumstellar disc edge-on (Hanuschik 1995). It has been emphasized by 
Rivinius, Stefl \& Baade (1999) that such features develop in the inner regions of 
circumstellar discs and are an indication of  gas transfer from the star to the disc. In 
HD~52721 the high rotational line broadening will prevent that such CQE's are observed in the 
weaker metal lines. For this star, however, an interpretation of the feature in terms of a tidal 
effect due to the nearby companion (see Sect. 4) should also be considered. This should be 
checked by taking spectra at different phases of the binary motion. 

   The He\,{\sc i} lines of HD~53367 have photospheric absorption profiles with an average 
FWHM of 150 km~s$^{-1}$. In the 5876~\AA\ line profiles (Fig. 4b) we observe the presence 
of an emission component, with a blue-shift of $\approx$ 150 km~s$^{-1}$ with respect to 
the line centre. The profile of Jan. 16, 1995 shows 
somewhat larger emission and deeper absorption components than that, taken 31 days
before. These components are also clearly present in the corresponding profiles for HD~53367 and
perhaps LkH$\alpha$~220 obtained by Finkenzeller \& Mundt (1984) in Nov. 1981. 
The emission and absorption
components of this line seem to vary in time (see Table 4). Inspection of the He\,{\sc i} 6678~\AA\ 
profiles of HD~53367 (Corporon \& Lagrange 1999) show variability on time-scales of less than 26 days, 
which could partly be due to radial velocity variations with a probable (binary) period of 166 days but 
changes in the width of this absorption profile suggest also here the presence of a variable 
emission component. Because the binary parameters suggest a large separation of the companions the
variations can not be due to tidal interactions. Since the emission feature in the profile of the 
5876~\AA\ line is not in the centre of the line we think it cannot be interpreted as a CQE, but the 
blue-shift of the emission suggests a hot plasma flowing from the photosphere to the inner 
circumstellar disc. We should add here that the profile of the photospheric He\,{\sc ii} 1640~\AA\ 
line also shows indications for the presence of blue-shifted ($\approx$ 150 km~s$^{-1}$) emission. 
The parameters of the profiles of the H$\alpha$, He\,{\sc i} (5876~\AA), 
Na\,{\sc i}\,D and Ca\,{\sc ii}\,(2) lines of HD~52721 and HD~53367 are collected in Table 4.

The profiles of several other lines in the spectrum of HD~52721 (H$\gamma$, 
He\,{\sc i}
(3888.6, 3964.7, 4437.9, 4471.5~\AA), H$\varepsilon$, and Ca\,{\sc ii}\,H and 
Ca\,{\sc ii}\,K) are shown in Fig.~6.  For H$\gamma$ we find an EW of
3.60 $\pm$ 0.06~\AA, or 3.96 $\pm$ 0.04~\AA, if we correct for a small 
contribution
of emission in the line centre. This can be compared with an EW of 3.0~\AA,
estimated from the spectrum of Finkenzeller \& Jankovics  obtained on Feb. 18, 
1981. The value of the EW of H$\gamma$ is consistent with the prediction of 
the Kurucz model for $T_{\rm eff}$ = 25,000~K and $\log g$ = 4.0 (B1\,IV), which
supports the classification of Table 2.
\begin{figure*}
\centerline{\psfig{figure=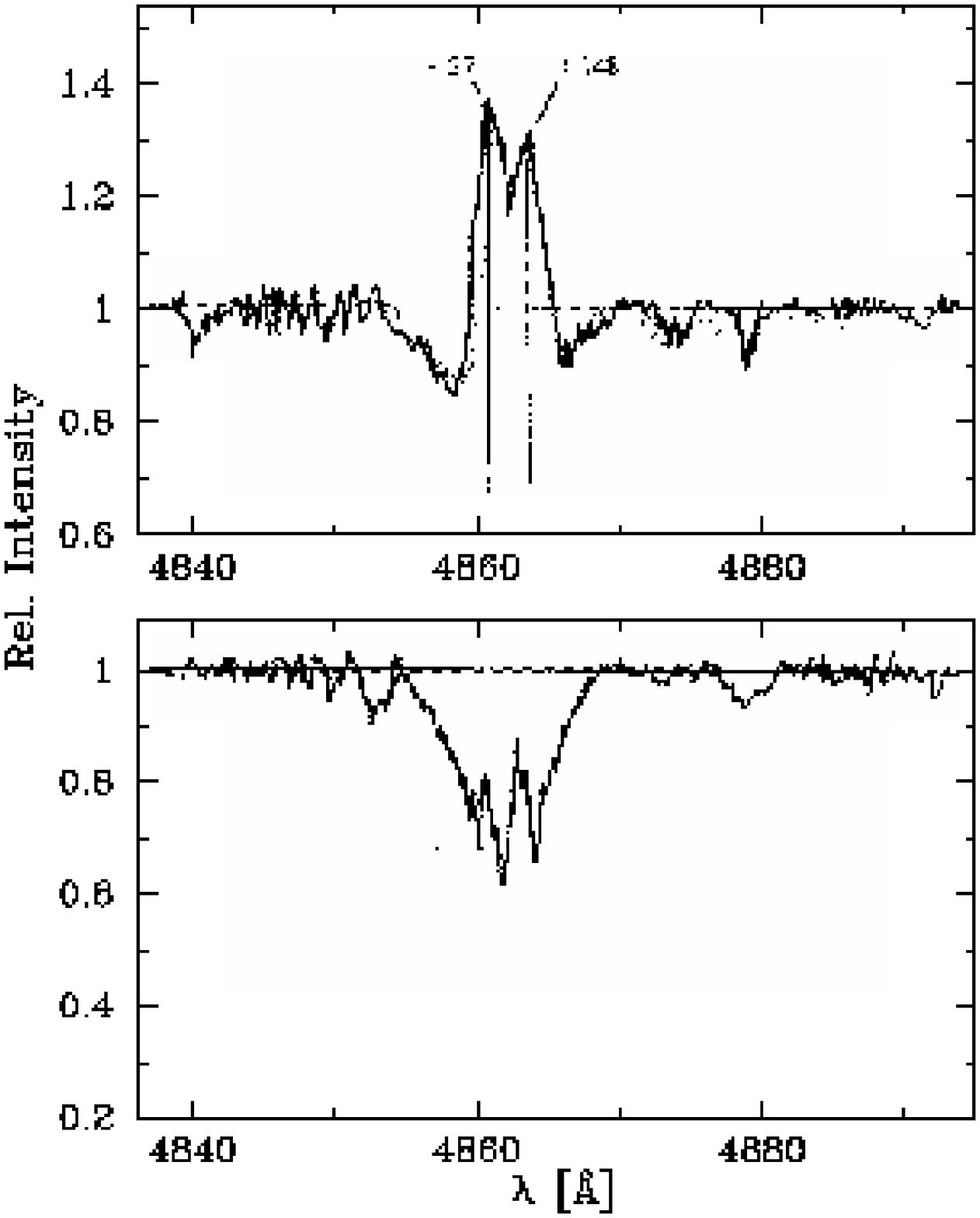,height=8.5cm,angle=0}
            \hspace*{0.3cm}
            \psfig{figure=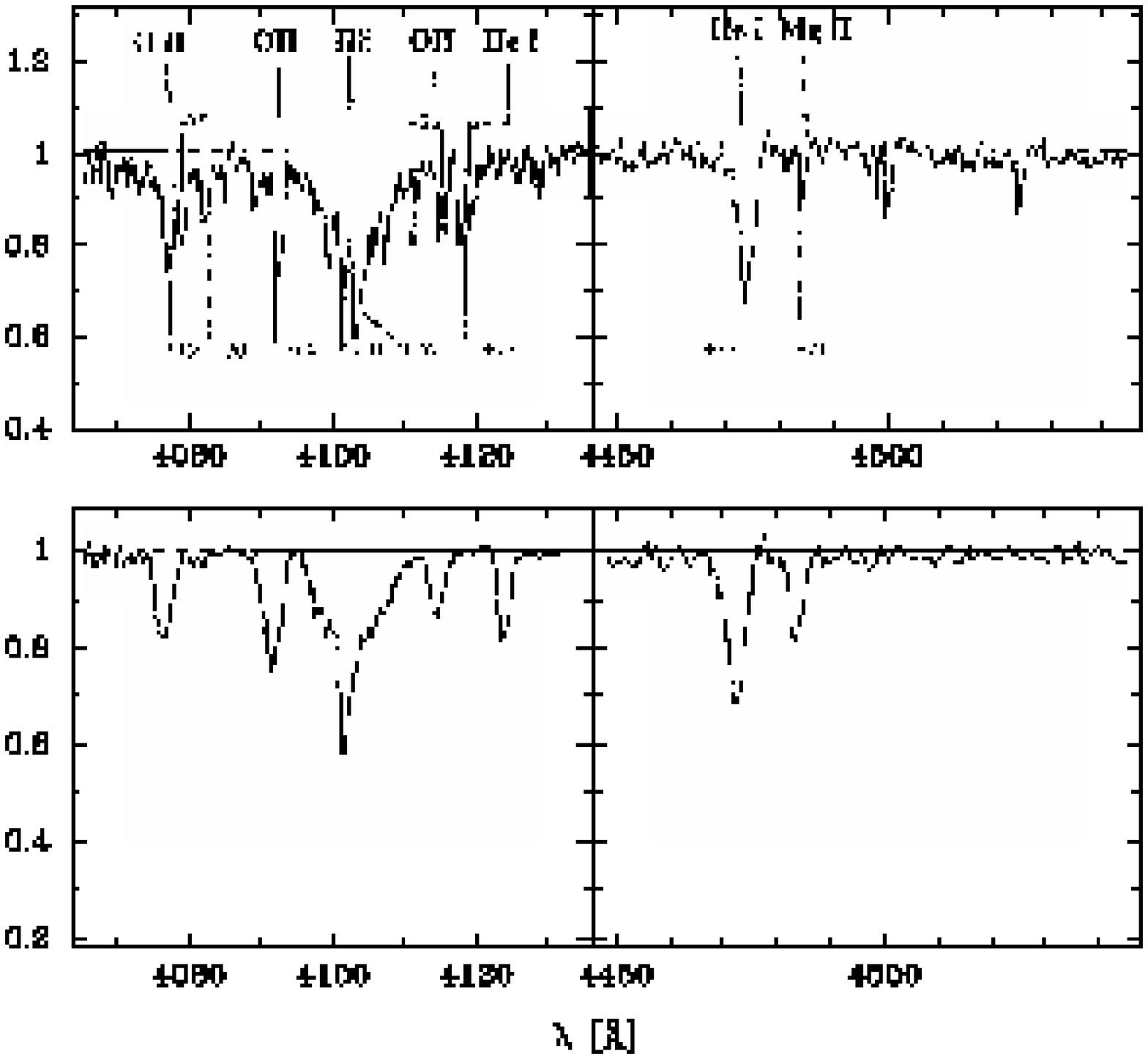,height=8.5cm,angle=0}}
\caption[]{High resolution spectral intervals of HD~53367 (spectra of the Main
 Spectrograph of the SAO).  The left panels show the profiles of H$\beta$,
 whereas the right panels display the profiles of H$\delta$, He\,{\sc i}
 (4120.8 and 4471.5~\AA) and Mg\,{\sc ii} 4481~\AA.  The panels 
 correspond with observation dates 
 02 March 1991, at maximum brightness (upper panels) and 15 Oct. 1989, at
 minimum brightness phase of the star (lower panels).}
\end{figure*}

    In Fig.~5 we show the low resolution spectra of the H$\alpha$ emission 
stars HD~52721,
HD~53367 and LkH$\alpha$~220. The latter star has the strongest emission in 
H$\alpha$ and H$\beta$,
but a good
comparison is difficult, since the three stars are variable in emission. The 
blue low resolution
spectra of HD~52721 and HD~53367 are similar to the spectra of these stars 
which were obtained
by Finkenzeller \& Jankovics (1984) with higher resolution. Their main 
absorption lines are those
of He\,{\sc i} as expected for these early B spectral types. Comparison of the 
higher resolution
blue 
spectrum of HD~53367 published by Finkenzeller \& Jankovics (1984) with that of 
the spectrum
of $\tau$~Sco (B0\,V) in the same paper, shows faint absorption lines of 
O\,{\sc ii}\,(2), (5), (6), (10)
and a faint line of Si\,{\sc iv} (1) at 4116~\AA, which also occur in the 
spectrum of $\tau$~Sco, and 
therefore probably are formed in the photosphere. 
The presence of such  faint lines cannot be verified in the spectrum of 
HD~52721, because of
the broadening of these lines in this fast-rotating star.
In contrast to the blue spectra of HD~52721 and HD~53367, that of 
LkH$\alpha$~220 (B5\,V)
shows emission in the photospheric Fe\,{\sc ii}\,(42) lines at 4923.9, 5018.4 
and 5169.0~\AA\
(B0 with 30,000~K is
too hot for Fe\,{\sc ii}). The He\,{\sc i} lines at 5876 and 6678~\AA\ and the 
Na\,{\sc i}\,D
lines are dominated by absorption, as well as the Ca\,{\sc i} lines at 4026 and 
4435~\AA\ and those
of probably Mg\,{\sc ii} (4481~\AA) and Ti\,{\sc ii} (4469~\AA).  There is a 
weak indication
of emission of [O\,{\sc i}] (6300~\AA), also observed by B\"ohm \& Catala 
(1995).
The low resolution blue spectra of LkH$\alpha$~220 and LkH$\alpha$~218, taken 
by
Cohen \& Kuhi (1979) on Jan. 14, 1977 show no clear lines except the Balmer and 
Na\,{\sc i}\,D
lines. 
In these spectra the H$\alpha$ emission of LkH$\alpha$~220 is stronger than 
that of
LkH$\alpha$~218.  
In the red spectrum of LkH$\alpha$~220 (Fig.~5) one can distinguish the 
7774~\AA\ and 8446~\AA\
lines of O\,{\sc i}\,(1) in emission. The EW of the unblended 7774~\AA\ line is 
$\approx$ $-$0.6~\AA, compared with 
+0.6~\AA\ for a standard B5\,IV--V star (Fig.~2 of Slettebak 1986). As for 
HD~52721 and HD~53367, the 
red triplet of Ca\,{\sc ii}\,(2) of LkH$\alpha$~220 is in emission and blended 
with emission from P13, 
P15 and P16 (see insert of Fig.~5). In this spectrum the emission flux of the 
unblended Paschen
lines is decreasing from a strong P11 emission at 8863~\AA\ to a very weak 
emission in the P17 line. 
\begin{figure}
\centerline{\psfig{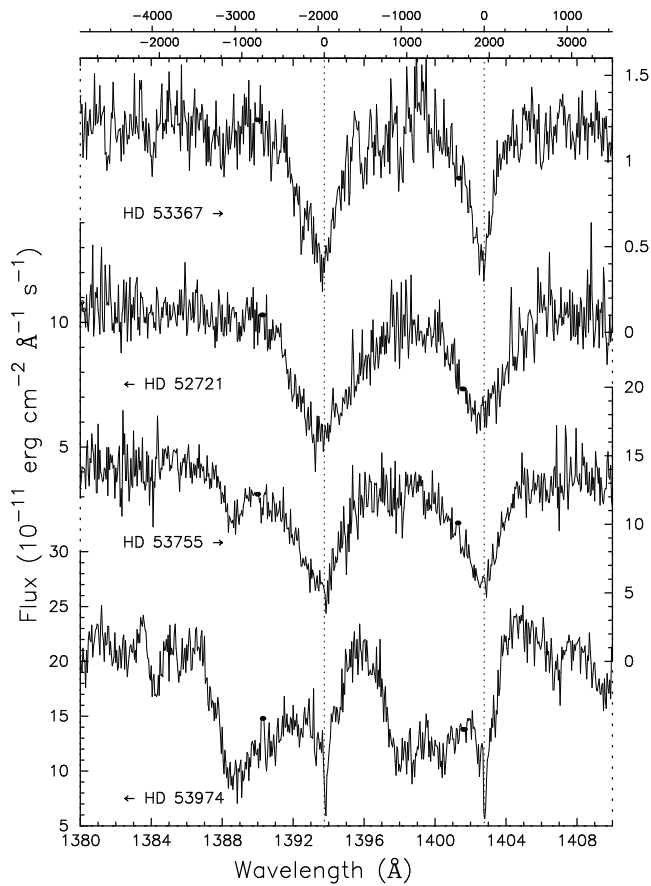}}
\caption[]{High resolution {\it IUE} profiles of the Si\,{\sc iv} resonance 
 lines for HD~52721, HD~53367, HD~53755 and HD~53974.  The arrows near 
 the stellar names point to the corresponding flux scales.}
\end{figure}

In the high resolution {\it IUE} spectra most of the UV lines are formed in the 
photospheres and the cooler parts of the inner circumstellar region. The 
profiles of the
Si\,{\sc iv} resonance lines have blue wings extending up to 530 km~s$^{-1}$ 
for HD~52721 and
up to 470 km~s$^{-1}$ for HD~53367 (Fig.~8). Those of the C\,{\sc iv} resonance 
lines extend
up to 340 km~s$^{-1}$ for HD~52721 and to 370 km~s$^{-1}$ for HD~53367. These 
lines are
formed in outflowing winds, accelerated in the circumstellar envelope.

\subsubsection{The B stars: HD~52942, HD~53623, HD~53755, HD~53974 and HD~54439}
   From the Walraven photometry (Th\'e, Wesselius \& Janssen 1986) we measured 
the fill-in of the
Balmer 
discontinuity of HD~53755 and HD~53974 and found values of $\Delta D_b$ of 0.15 
and 0.20 mag
respectively. The lack of H$\alpha$ emission in the former two stars must 
therefore be due
to a much lower $T_{\rm shell}$ ($<$ 10,000~K) and consequently a much lower 
emission measure
for the inner circumstellar envelope of the stars.  This temperature may be too 
low to populate
the upper level of the Balmer series.

   For the five B stars, in which H$\alpha$ is in absorption, there is a 
certain
similarity in profile shape between H$\alpha$ (Fig.~9a) and He\,{\sc i} 
(5867~\AA) (Fig.~9b).
This suggests that H$\alpha$, similar to He\,{\sc i} (5867~\AA), is formed in 
the
photosphere of these stars. The profile parameters (EW and FWHM) of H$\alpha$ 
and He\,{\sc i} for
the five stars are given in Table 6. A comparison with the EWs of H$\alpha$ 
predicted
by Kurucz models shows a general agreement with the UV spectral 
classification of Table 2.  In the spectrum of HD~53623 both lines show a 
narrow, deep
and symmetrical central absorption core. Since this object is a single star, it 
suggests that
this central component is formed in a shell. The FWHM of the shell component of 
H$\alpha$
is $\approx$ 40~km~s$^{-1}$ and that of the He\,{\sc i} line is $\approx$ 
60~km~s$^{-1}$.

Apart from this shell component in the H$\alpha$ and He\,{\sc i} profiles of 
HD~53623 there is a
similarity between the He\,{\sc i} profiles of HD~53623, HD~53755 and HD~52721 
and also
between the H$\alpha$ profiles of the first two stars. This is especially 
expressed by the
similarity
in their half-widths and probably also by their values of $v \sin i$. If we lay 
the corresponding
profiles on top of each other we observe differences in central depth. Since 
the spectral types of
these stars are comparable it is conceivable that (similar to HD~52721) the 
H$\alpha$ and
He\,{\sc i} 
profiles of HD~53623 have emission components with respect to those of HD~53755.
There is also some similarity between the line profiles of HD~53974 (FN CMa) 
and HD~54439,
but the centre of the He\,{\sc i} line of HD~53974 is deeper than that of 
HD~54439, which could be
due to a difference in  luminosity type or in the contribution of an emission 
component for
HD~54439. We shall return to this point in the discussion (Sect. 5). 

    The H$\alpha$ profile of HD~52942 (FZ CMa) is very broad and has a core 
which can either be
double or filled in by emission. The double nature of the profile seems most 
probable, since
the object is a double-line eclipsing binary with a period of 1.27 days (Moffat 
\& Vogt 1974)
with a maximum radial velocity separation of 342 km~s$^{-1}$ (at phase 0.25 and 
0.75). From the
ephemeris of Moffat \& Vogt we conclude that our spectrum was taken at phase 
0.635 from the
primary minimum. 
The separation of the H$\alpha$ components in our spectrum is $\approx$ 200 
km~s$^{-1}$,
which is compatible with this phase. 

Most of the UV lines are formed in the photospheres and cool circumstellar 
regions of these stars.  The profiles of the Si\,{\sc iv} and C\,{\sc iv} 
resonance lines are asymmetrical and their blue wings extend to velocities of 
$\sim$ 1600 km~s$^{-1}$ (Fig.~8; for HD~53974 see also Friend, 1990). This 
gives evidence for fast outflowing winds from these stars.  
Accretion discs can not accelerate the winds to such high velocities 
so that there must be a different driving mechanism for the winds 
of these stars.

\section{Variability}
\begin{figure*}
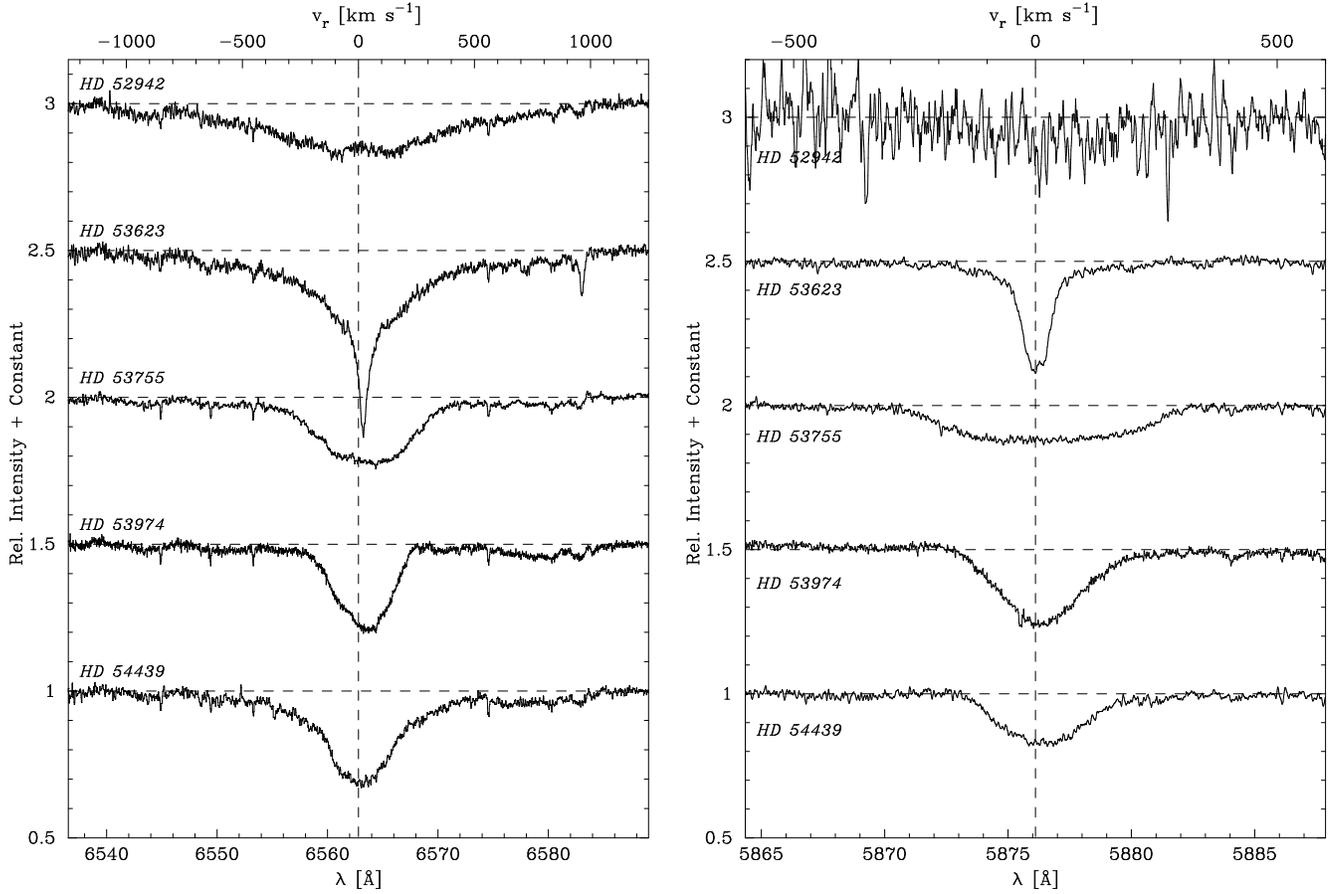

\centerline{\psfig{figure=fig9a.ps,width=8.5cm,angle=0}
\hspace*{0.3cm}
\psfig{figure=fig9b.ps,width=8.5cm,angle=0}}
\caption[]{High resolution H$\alpha$ (left) and He\,{\sc i} 5876~\AA\ (right) 
profiles
 of HD~53623, HD~53755, HD~53974 and HD~54439.}
\end{figure*}
\begin{table*}
\centering
\caption{H$\alpha$ and He\,{\sc i} Equivalent Width (EW) and Full Width Half 
Max. (FWHM) for
 five absorption-line stars in CMa R1.}
\small
\begin{tabular}{@{}llclccc}
\hline
Line                    & HD 53755         &	HD 53974          &  HD 54439      
  &  HD 53623         &  HD 52942\\
\hline
H$\alpha$		& 		   &			  & 		     & 		         & \\
EW   [\AA]              & 3.02 $\pm$ 0.11  &	2.54 $\pm$ 0.18   &  3.56 $\pm$ 
0.10 &  4.75 $\pm$ 0.06  & 3.71 $\pm$ 0.12\\
EW*  [\AA]              & 3.21 $\pm$ 0.09  &			  &  3.70 $\pm$ 0.19 & 	   	     
    & \\
EW** [\AA]              & 		   &			  & 		     &  4.34 $\pm$ 0.12  & \\
FWHM  [km~s$^{-1}$]     & 480              &	280               &  320           
  &  460              & 807 \\
FWHM* [km~s$^{-1}$]     & 420              &			  &  285             & 		        
 & \\
$T_{\rm eff}$-mod.      & 25,000           &	30,000            &  25,000        
  &  25,000           & 20,000\\
$\log g$-mod.           & 3.9              &	4.0               &  4.14          
  &  4.43             & 3.59\\
Sp. class               & B1\,IV            &	B0\,IV            &  B1\,IV/V       
  &  B1\,V             & B2\,IV\\
\\
He\,{\sc i} (5875.6~\AA)& 		   &		          & 		     & 		         & \\
EW  [\AA]               & 1.03 $\pm$ 0.10  &	1.06 $\pm$ 0.07   &  0.76 $\pm$ 
0.05 &  0.84 $\pm$ 0.05  & --\\
EW* [\AA]               & 1.22 $\pm$ 0.07  &			  & 		     & 		         & \\
EW** [\AA]              & 		   &			  & 		     &  0.37 $\pm$ 0.06  & \\
FWHM  [km~s$^{-1}$]     & 400              &	200               &  210           
  &  250:		 & \\
FWHM* [km~s$^{-1}$]     & 435              &                      &             
     &                   & \\
\hline
\end{tabular}
\noindent
\flushleft
EW* = EW after estimated correction for emission component. EW** = EW without 
``shell''
contribution (for HD 53623). For H$\alpha$ the corrected EWs are compared with 
the predictions of
the models of Kurucz (1991) which lead to interpolated values of $\log g$ and 
therefore serve to check the luminosity classes of Table 2. 
\end{table*}
 We shall discuss here the photometric variability of the four Be stars 
and some first attempts to interpret the variations.

\subsection{HD~52721 (GU CMa)}
\begin{figure*}
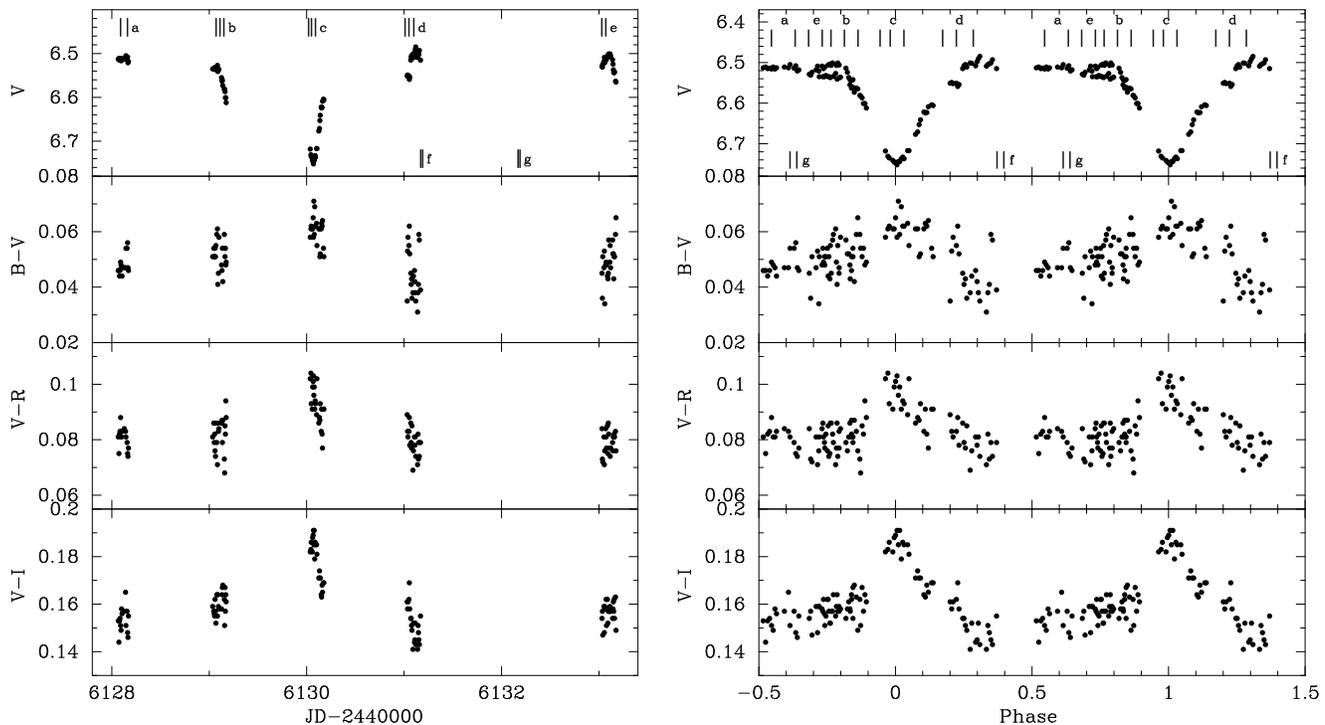

\centerline{\psfig{figure=fig10a.ps,height=9.5cm,angle=0}
\hspace*{0.3cm}
\psfig{figure=fig10b.ps,height=9.5cm,angle=0}}
\caption[]{Light curve (left) and phase diagram (right) of HD~52721 in $BVRI$ 
 during 5 nights in March 1985.  For the construction of the phase diagram a 
 period of 0.80508 days was assumed.  Vertical marks indicate the times at 
which
 H$\alpha$ (top) and Mg\,{\sc ii} 2800~\AA\ (bottom) data were obtained.}
\end{figure*}
This star is a visual binary (ADS 5713) with a separation of 0.65\arcsec\ 
and $\Delta m$ = 0.95 mag. 
The Hipparcos photometry (ESA 1997) is consistent with a period of 1.61037 
days. During the period
1987--1998 extensive photometry (320 observations over 12 seasons) of this star 
was collected at
Mt. Maidanak (Uzbekistan) in the framework of the ROTOR photometric program. 
The light-curve
over 10 years does not show clear ($>$ 0.1 mag) indications of long-term 
variations
of upper and lower limits in $V$.  The results of a period 
search on these data has recently been published (Ezhkova 1999). Two periods 
with equal probability
were found: 1.610158 and 0.80508 days. For the shortest period the date of the 
minimum is given by:
$JD_{\rm min}$ = 2447018.115 + 0.80508 $\times$ E days. 
Earlier monitoring of HD~52721 in $UBVRI$ during 5 nights (107 observations) in 
1985
(Praderie et al. 
1991) appear to be in excellent agreement with these results (Fig.~10). From 
the photometric
data of 1985 we derived the extinction corrected $UBVRI$ fluxes at maximum and 
minimum brightness and
compared the variations with the corresponding intrinsic fluxes of the Kurucz 
model.  It appears that
the flux variations from maximum to minimum brightness correspond roughly  with 
a change
in $T_{\rm eff}$ from 25,000~K (B1\,V) to 22,500~K (B1.5\,V). The simultaneously 
obtained high
resolution H$\alpha$ profiles (at phases a, b, c, d and e in Fig.~10) show no 
difference in
shape but a 40\% increase in EW at photospheric minimum c (Praderie et al. 
1991).  Several interpretations of this periodicity may be considered:

{\it a. An eclipsing binary system.} 
The light curve for the longest period is consistent with a binary system with 
two components with
equal radii and effective temperatures. However, the radial velocities of the 
He\,{\sc i} 5875\AA\
line in a sequence of 0.7~\AA\ resolution spectra of HD~52721, taken in 1991 
over a period of several weeks with the main
spectrograph on the 6-m SAO telescope, show variations with an amplitude not 
larger than $\approx$ 20 km~s$^{-1}$, which is 10--15 times smaller than 
expected for such a system with components of equal mass (5--10~M$_\odot$).  
The radial velocity variations in the profiles of He\,{\sc I} 6678~\AA,
observed by Corporon \& Lagrange (1999) are consistent with the SAO data.
However, if the mass of the secondary is at least five times smaller 
than that of the primary component, 
an upper limit of $\approx$ 30 km~s$^{-1}$ is not in conflict with the shorter 
period of 0.80508 days.  In this case we do not observe a secondary minimum 
in the visual light curve because the eclipse is only partial.  Further 
support for the hypothesis that we are dealing with a secondary that is much 
less massive than the primary comes from the observation that the colours 
of the system become slightly redder at minimum brightness (Fig.~10).
With a distance of 1050~pc and $T_{\rm eff}$ = 25,000~K we derive a 
luminosity of $1.8\times 10^{4}$~L$_\odot$ and a radius of 7.1~R$_\odot$ 
for HD~52721. From theoretical
evolutionary models (Schaller et al. 1992) we then estimate that the mass of 
the primary should be close to 12~M$_\odot$. The orbital distance is then 
$\approx$ 9~R$_\odot$, which is large enough to accommodate a 1~M$_\odot$ 
secondary star. 
We estimate that the secondary could be a K0 star (if its age 
is not much larger then that of the primary).

{\it b. A pulsation instability.}
In the HR diagram HD~52721 is situated close to the predicted border of the 
$\beta$ Cep instability
zone for masses 10--12 M$_\odot$ (Pamyatnykh 1999). The pulsation periods 
predicted by the model
depend on various input surface parameters such as hydrogen and metal 
abundances and
rotation. For a mass of 12~M$_\odot$ and $\log T$ = 4.35 (B1.5\,V), solar 
abundances and no rotation,
pulsations with periods between $\approx$ 0.4 and 1.6--2.0 days are possible in 
the second order mode ($l$ = 2).
For $\log T$ = 4.3 (B2\,V), the period can range from 0.8 up to 2.4 days. Both 
measured periods are in this range. 
It is not clear if the magnitude of the photometric and line profile (H$\alpha$ 
and He\,{\sc i}) variations can be explained by these pulsation models.

\subsection{HD~53367 (V750 Mon)}
HD~53367 is a visual binary with a separation of 0\farcs65, a magnitude 
difference of 1.41 mag and a
position angle of 298\degr\ (ESA 1997).  HD~53367 is also a spectroscopic 
binary (Finkenzeller \&
Mundt 1984). The main component shows radial velocity variations in the 
He\,{\sc i} 4471, 5876 and 6678~\AA\ lines, which can be fitted with a 
period of 166 days and an amplitude 
of about 20 km~s$^{-1}$ around an average velocity of +48.2 km~s$^{-1}$.  
A possible orbital solution
has $e$ = 0.18 and $\omega$ = 303\degr\ (Corporon \& Lagrange 1999).
  
HD~53367 shows a low $v \sin i$ (20--40 km~s$^{-1}$).
There are several indications that this is not due to a small inclination, 
i.e. that we are not observing the star pole-on. 
One indication is the observation by Oudmaijer \& Drew (1999) of the variation 
of the linear
polarization percentage over the H$\alpha$ profile of HD~53367, which suggests 
that the ionized
envelope, projected on the sky, is not observed as a circular disc as expected 
in case the star is observed pole-on. 
 Another indication against a pole-on view of HD~53367 has been given by
Oegerle et al. (1983). These authors argued that the simultaneous occurrence of 
emission in the
8446~\AA\ O\,{\sc i} line and absorption in the 1302~\AA\ O\,{\sc i} line can 
only be understood if
the star has an envelope, which is observed at large inclination with the polar 
axis (see also Sect. 3.3.1).  

Apart from its velocity variations, HD~53367 also shows photometric variability.
Hipparcos observed an amplitude of 0.24 mag, but no period is given (ESA 1997). 
However, over a longer interval than these Hipparcos observations, the 
photometry of the ROTOR group, the Wesleyan Observatory and the Corralitos 
Observatory suggests
a mean periodic variation with a period of about 9 years with an amplitude of 
about 0.3 mag (Fig.~11).  The colour-magnitude diagram of HD~53367 shows that 
the colours are bluer at minimum brightness. Nevertheless after correction of
the $UBVR$ fluxes for extinction it appears that in the minimum phase the 
photospheric temperature has decreased from 30,000~K to about 25,000~K.

At minimum brightness (Oct. 15, 1989) and at maximum brightness (March 2, 1991) 
high resolution spectra have been taken with the Main Stellar  Spectrograph on 
the 6-m BTA telescope of the SAO (Fig.~7).  They show that in the minimum 
phase the emission component in all the lines is weaker than in the maximum 
brightness phase.  
The emission of H$\beta$ at minimum brightness is very faint. This could
be related with variations in the H$\alpha$ profile, reported by Halbedel 
(1989), but the time-scale of these profile variations is not known yet, 
except for small changes in H$\alpha$ after 1 month (Fig.~4a).  
H$\alpha$ profiles of May--Dec. 1974 (Garrison \& Anderson 1977), July 1981 
(Finkenzeller \& Mundt 1984), 1983 (Andrillat 1983), June 1986 (Hamann \& 
Persson 1992), Nov. 1987 (Halbedel 1989) and our observations of Dec. 1994 
and Jan. 1995 (Fig.~4a) show differences, although they are small. However, 
the EWs of H$\alpha$ show a strong decrease between Dec. 1987 and Jan. 1992 
(Table 4), a period for which no H$\alpha$ line profile information is available.  
For H$\delta$ (Fig.~5), the contribution of emission to the profile is small. 
We find an EW of 3.31 $\pm$ 0.08~\AA\ for H$\delta$ in 1991 and 3.73 $\pm$ 
0.16~\AA\ for this line in 1989. These values are consistent with the 
predictions of Kurucz models for $T_{\rm eff}$ = 30,000~K, $\log g$ = 4.25 
(B0\,IV) or 25,000~K and $\log g$ = 3.75 (B0\,III/IV).
\begin{figure}
\centerline{\psfig{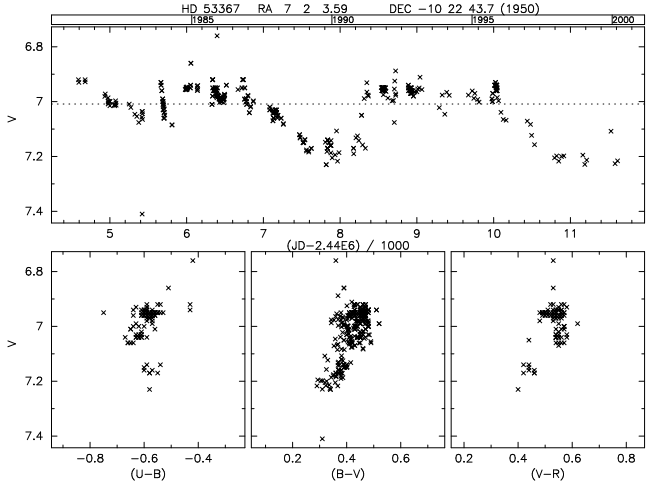}}
\caption[]{Light curve (top) and colour-magnitude (bottom) diagrams 
 of HD~53367.}
\end{figure}

It may be noted here that HD~53367 is not the only young Be star with a long 
photometric period. Another early type Herbig Be star, BD+65\degr1637 shows a 
similar behaviour, with a period of at least 15 years. It is to be expected 
that more long periodicity young variables will be detected when the 
photometric observations are prolonged.
We can think of several possible causes for such long photometric periods:

{\it a. Magnetic activity of a B star.}
\noindent
Little is known about the magnetic fields of early type B-stars. 
The intrinsic continuum polarization HD~53367, observed by 
Oudmaijer \& Drew (1999), is
very low and has an angle of 44 degrees, which agrees with that of the 
dust arc in CMa R1 (Vrba, Baierlein \& Herbst 1987).  This suggests 
that the rotation axis of  HD~53367 ``remembers''
the larger scale circumstellar field direction, which means that at 
least at the time of formation of the dust arc, the star had a magnetic 
field that could be coupled with the large scale field of CMa R1. 
The exceptionally low rotation velocity of HD~53367 may therefore be
a result of magnetic braking during outflow or accretion. 

Attempts to measure the magnetic field of HD~53367 have been made recently 
by Glagolevskij \& Chountonov (1998). Although no significant field was detected 
at that moment it does not exclude the possibility of the presence of weak 
surface fields at other moments.
We suggest that every 9 years during a short period the internal field
moves upward and emerges in the surface layers of the star, thereby 
producing spot formation,
similar to the solar activity cycle. If the spots cover a large part of the 
surface they may lower
the effective temperature during the activity phase and may have an effect on 
the photometric
colours. If the surface magnetic field was stronger during the earlier 
evolutionary phases of the star (e.g. during the convective phase) the field
could also have braked the rotation of the star during outflow or accretion. 

   The braking time due to angular momentum loss in  outflow has 
been calculated by
Friend \& MacGregor (1984).  If the Alfv\'en radius $R_a$ of the magnetosphere 
is larger than the
corotation radius $R_c$ the matter trapped in the magnetosphere at $R_a$ will 
lose angular momentum
and will brake the rotation of the star.  For a 20~M$_\odot$ star with a 
rotation velocity
of 250 km~s$^{-1}$ and a 200~G surface field we find an Alfv\'en radius of 
1.74~R$_\star$.
The outflow velocity at $R_a$ is then $\sim$ 700 km~s$^{-1}$. From equation 38 
of Friend \& MacGregor's paper it is then seen that a mass outflow 
of $4 \times 10^{-6}$~M$_\odot$~yr$^{-1}$ in a field of only 200~G is able to 
brake the rotation of the star within 0.6~Myr.  Similar results are obtained 
if one estimates the braking during the pre-main sequence
accretion phase (Rathnasree 1994). At an accretion rate of 
$4 \times 10^{-6}$~M$_\odot$~yr$^{-1}$, a field of 400~G will be large enough 
to brake the rotation from close to break-up velocity to very slow 
rotation in about 1.7~Myr. In order to arrive at the same rotation velocity 
after a time-scale of 6~Myr smaller accretion rates and smaller fields would 
be sufficient. 
 
{\it b. A rotationally distorted Be star.} 
The long-period photometric variation may be due to a quadrupole contribution 
in the external potential of a rotationally
distorted Be star (Papaloizou, Savonije \& Henrichs 1992). We should then 
expect a simultaneous periodic
variation in the V/R ratio of the Balmer emission profiles. We have noted a 
change from V/R $<$ 1 for the profile observed in 1974 to  values of 
V/R $>$ 1 for the profiles obtained in 1981 and 1994,
but we do not have enough high resolution H$\alpha$ observations to detect a 
periodicity.
\begin{figure}
\centerline{\psfig{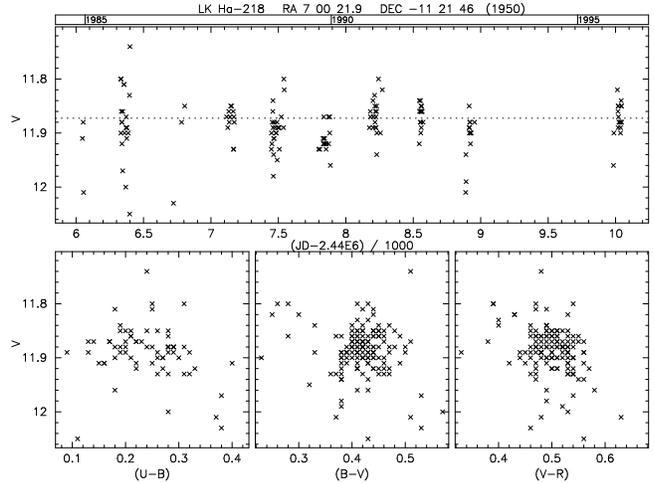}}
\caption[]{Light curve (top) and colour-magnitude (bottom) diagrams 
 of LkH$\alpha$~218.}
\end{figure}
\begin{figure}
\centerline{\psfig{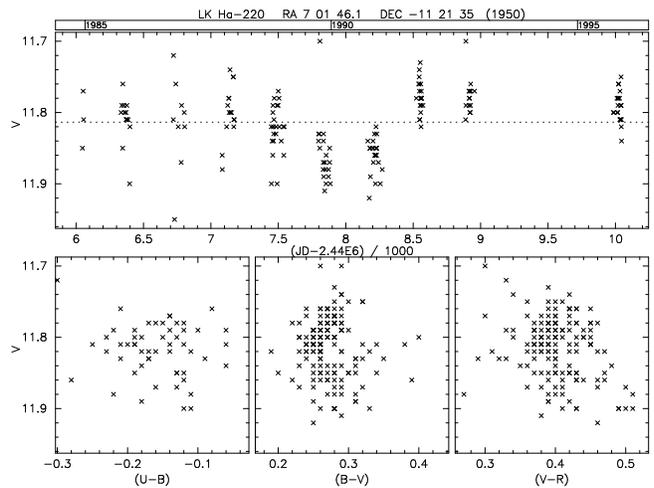}}
\caption[]{Light curve (top) and colour-magnitude (bottom) diagrams 
 of LkH$\alpha$~220.}
\end{figure}

\subsection{LkH$\alpha$ 220 (HU CMa) \& LkH$\alpha$ 218 (HT CMa)}
   In the course of the $UBVR$ photometric monitoring program (ROTOR) 
at Mt. Maidanak, 153 observations of LkH$\alpha$~218 and 
168 observations of LkH$\alpha$~220 were made between 1985
and 1995. The minimum time interval between the observations is 0.8 days. 
The variability was
classified by Shevchenko et al. (1993) as LQ and BF respectively. Both stars 
vary irregularly
with low amplitude ($\la$ 0.07 mag) on short time scales (2 days) but around 
1990 the lightcurve
of LkH$\alpha$~220 also shows a broad and shallow  minimum during about 600 
days (Fig.~13). For both
stars there were a few variations with larger amplitude (0.2--0.4 mag). These 
variations were characterised by reddening during a decreasing brightness, 
as expected from an increasing circumstellar dust extinction. The slope 
of these variations in the colour-magnitude diagram
is close to 3. The presence of circumstellar dust is also evident from 
the SEDs (Fig.~1).

\section{Summary and Discussion}
In the previous sections we have studied the photometric properties and the 
high resolution
profiles of  several lines in the spectra of the early B-type stars HD~52721, 
HD~53367, LkH$\alpha$~220,
LkH$\alpha$~218, HD~52942, HD~53623, HD~53755, HD~53974 and HD~54439, which are 
members of
CMa R1 (paper I). The membership is confirmed by the common foreground excess.
We  have described the photometric behaviour of the first four stars and we
have discussed the possible origin. 
The conclusions of this discussion can be summarized by: (a) HD~52721 varies
with a period of 0.80508 days and the variation is consistent with
the presence of a partially eclipsing binary companion. (b) HD~53367 has a 
long-term photometric
variation with a period of $\approx$ 9 yr, and a shorter variation (166 days) 
in radial velocity. The
second variation (detected by Corporon \& Lagrange 1999) is probably due to the 
presence
of a binary companion. We have given arguments for the interpretation of the 
long-term
photometric variability in terms of a cycle of magnetic activity. During its 
previous evolution,
an outflow of $5 \times 10^{-7}$~M$_\odot$~yr$^{-1}$ in the presence of a 
modest (200~G)
magnetic field would have been sufficient to brake the rotation of the star 
to its present low value of
$v \sin i$. (c) The photometric variability of LkH$\alpha$~218 and LkH$\alpha$~220 
seems partly due to
variability of the column density in their circumstellar dust envelopes.

Three of these stars (HD~52721, HD~53367 and LkH$\alpha$~220) have spectra, 
which are characterised
by emission components in H$\alpha$, H$\beta$, He\,{\sc i} (5867~\AA), the 
Ca\,{\sc ii}\,(2) red triplet,
the O\,{\sc i} 7773~\AA\ triplet 
and the O\,{\sc i} 8446 line, which gives arguments for the presence of a disc 
around these stars (see
also Hamann \& Persson 1992). Whether these discs are classical Be star discs 
or the remnants
of pre-main sequence accretion discs remains uncertain so far. Note that 
stellar winds from early
B type main sequence stars and classical Be stars usually reach velocities in 
excess of 1500 km~s$^{-1}$
(see e.g. Porter 1999), much higher than  the 600 and 450 km~s$^{-1}$ as 
observed for HD~52721
and HD~53367 in the wings of the Si\,{\sc iv} lines (Fig.~8) and the 340 and 
370 km~s$^{-1}$ in the
wing of the 1548~\AA\ C\,{\sc iv} line.  However, recently Drew, Proga \& Stone 
(1998) modelled the wind from
accretion discs of young early type B stars and found that rotating discs can 
have high mass-loss rates
($>$ 10$^{-7}$ M$_\odot$~yr$^{-1}$)  in combination with lower velocity (a few 
hundred km~s$^{-1}$)
equatorial disc winds.  These are precisely the mass losses and wind velocities 
we measured for
HD~52721 and HD~53367.

    We also discussed the H$\alpha$ and 5876~\AA\ He\,{\sc i} profiles of the 
five stars without clear indications of
emission in H$\alpha$. In these stars H$\alpha$ and He\,{\sc i} have very 
similar profiles, which suggests that
H$\alpha$ is mainly formed in the photospheres of these stars. The high wind 
velocities measured in the wings
of their Si\,{\sc iv} profiles ($\approx$ 1500 km~s$^{-1}$) of HD~53755 
and HD~53974 are typical for normal early type stars.
If the four H$\alpha$ emission stars indeed have accretion discs, the
next question is whether the other five young stars HD~52942 (B2\,IVn), HD~53623 
(B0.5--1\,IV/V),
HD~53755 (B0\,V + F5\,III), HD~53974 (B0\,IIIn) and HD~54439 (B1\,V) could have 
such discs.  Although HD~53974,
HD~53755 and HD~54439 show weak indications of emission in their H$\alpha$ and 
He\,{\sc i} profiles
(Fig.~9), it seems doubtful if discs are present around these stars.

     One obvious question is then why these stars do not have accretion discs 
in spite of their ages being comparable with those of the stars with clear 
H$\alpha$ emission (paper I).  The conditions
for the existence of discs could be influenced by the presence of nearby binary 
companions.  Both
HD~52721 and HD~53367 have such companions, but of the other group also 
HD~53974 and
HD~52942 have companions at comparable distances. We suggest that the survival 
of an accretion
disc depends on influences from outside (winds or radiation from O stars or 
shock waves from
nearby supernovae) which could evaporate accretion discs around young, 
early-type B stars (Yorke \& Welz 1996).  
A natural candidate for such an outside disturbing source is the supernova 
about 0.8~Myr ago,
which accelerated the star formation in CMa R1 (Herbst \& Assousa 1978; see 
also paper I).

A strong indication for the position of this supernova is given by the centre 
of curvature
of the ionization front S296, which is near $\alpha$ (1950) = 7$^{\rm 
h}$08$^{\rm m}$,
$\delta$ (1950) = $-$11.2\degr\ (Reynolds \& Ogden 1978) and close to the 
original position of the runaway star HD~54662 (O7\,III), deduced from its 
motion in space
since the SN explosion 0.8~Myr ago. The other runaway candidate, 
HD~57682 (O9\,Ve; Comer\'on et al. 1998), passes the centre at a larger 
distance (if the Hipparcos proper motions are assumed).
It is remarkable that within the circle of curvature with radius 1\fd4 
($\approx$ 25.7~pc)
around the SN position only one star with H$\alpha$ emission, 
HD~55135 (B4\,Ve) (near the border and a foreground star at a distance 
of $\approx$ 500~pc with an $E(B-V)$ of 0.10 mag)
is present.  This is in sharp contrast to the field at the west side of S296, 
where in a dust cloud
region mapped with {\it IRAS} (see paper I and Luo 1991), the nine stars of our 
study are located.
In the field  between 6$^{\rm h}$50$^{\rm m}$--7$^{\rm h}$02$^{\rm m}$ and 
between 8\degr30\arcmin--12\degr00\arcmin, bordering S296 at its 
west side, Tovmasyan et al. (1993) have found 23 stars at a distance of 1100~pc 
from which six stars have H$\alpha$ emission. If we add LkH$\alpha$~218, 
LkH$\alpha$~220 and Z CMa,
we have 26 stars in this field from which nine stars have H$\alpha$ emission.  
The remarkable
contrast between the relative numbers of H$\alpha$ emission line stars 
east and west of S296 suggests that the dust clouds near this ionization
front shields the stars westward of S296 from UV irradiation by the SN or by 
the two O7 stars in the circle (HD~54662 and HD~53975).    

Recently several investigations have been devoted to the photo-evaporation of 
proplyds (discs)
during  FUV and EUV irradiation  by an external star (Johnstone, Hollenbach \& 
Bally 1998; Richling \& Yorke
1998; St\"orzer \& Hollenbach 1999). Until now few numerical estimates have 
been made and
mostly for low mass stars.  However, Richling \& Yorke (1998) have also 
computed the evolution
of a 1.67~M$_\odot$ disc around a massive (8.4~M$_\odot$) star as a result of 
irradiation from
outside with a flux of $10^{12}$ EUV photons~cm$^{-2}$~s$^{-1}$ (case A, model 
I) during 3000~yr.
The authors do not give estimates for the H$\alpha$ emission of this disc, as 
they did for
the low mass discs, but although the 1.67~M$_\odot$ disc is strongly deformed 
by the EUV
influx, it seems to have conserved its inner structure after the period of 
3000~yrs.
In the CMa R1 situation the O stars have an EUV photon flux of $1.5 \times 
10^{49}$~s$^{-1}$ (Cruz-Gonz\'alez
et al. 1974) at most, which for a distance of 18 pc to the dust barrier implies 
an upper limit of
$3 \times 10^8$  EUV photons~cm$^{-2}$~s$^{-1}$ at the location of the barrier
during perhaps 6~Myr. In comparison with
the numerical simulation of Richling \& Yorke these fluxes seem to be too low 
to affect the
discs of the stars east of the dust shield. The presence of several H$\alpha$ 
emission stars close to
HD~54662 but outside the circle (HD~55439, HD~54858, MWC~551) seems to confirm 
this.
The situation is quite different for the supernova. From a typical SN type II 
such as
SN 1979c in M100 (Panagia et al. 1980)  a flux of $\approx$ 300~ergs or 
$3 \times 10^{13}$ FUV photons~cm$^2$~s$^{-1}$ was flowing through a sphere at 
a distance
of 25.7~pc at the seventh day after the explosion. Since the decay of the 
luminosity is exponential, the initial flux may easily have been higher 
by one order of magnitude.  For other type II
supernovae such as the Crab SN and SN 1181 (Panagia \& Weiler 1980), the 
estimated FUV fluxes at 
25.6~pc are of the same order. Such fluxes may initially be 300--1000 times 
larger than assumed
in the simulation of Richling \& Yorke, but after 200 days  may have decreased 
to $3 \times 10^{10}$
FUV photons~cm$^{-2}$s$^{-1}$ (if the decay is similar to that of SN 1987A 
(Cassatella et al. 1987),
which is not a typical SN type II). Although the disc simulations for UV fluxes 
with such
time-behaviour have not been made yet, we expect an increased erosion rate of 
the outer disc
(emission) layers of stars in the neighbourhood of the ionization front. 
However, discs of stars westward of this front are shielded by extinction of 
the dust cloud.

Although these cloud extinctions in the direction perpendicular to the line of 
sight
are unknown, we can make an estimate of these from the 100~$\mu$m optical depth 
distribution
in the dust cloud, derived by Luo (1991) from the {\it IRAS} fluxes.  We assume 
that the
stars are in the mid-plane of the dust cloud at a geometrical depth of 15~pc 
(the extent of
the cloud on the plane of the sky is $\approx$ 30~pc) and extrapolate the 
optical depth
to the light path from the SN to the star.  To this cloud extinction in the 
light path we
add the circumstellar extinction, determined in the line of sight to the star.  
This
total absorption in the light path from the SN to each star gives the following 
percentages of transmission of the FUV (at $\approx$ 1500~\AA) flux: 12\% 
(HD~52721),
4.5\% (HD~52942), 1\% (HD~53367), 1.2\% (LkH$\alpha$~218), 6\% 
(LkH$\alpha$~220),
60\% (HD~53623), 40\% (HD~53755), 26\% (HD~53974), 28\% (HD~53456), 40\% 
(HD~53691), $<$ 10$^{-4}$\% (Z~CMa).  

This result shows that the stars with emission discs have received the 
lowest FUV photon fluxes from the SN, whereas the stars with high FUV
transmissions show no evidence for the presence of such discs.  The fact that 
HD~52942
shows no evidence for an emission disc in spite of its low incident FUV flux 
may be
related to the binary nature of this object.  We conclude that the presence of 
emission
discs in and around CMa~R1 appears to be correlated with the degree of 
shielding of the
discs against photons from a central supernova source.  A full explanation of 
emission
discs in the CMa OB1 association will require numerical simulations of disc 
evaporation
by the FUV photon flux from the supernova.  However, other effects of the SN 
explosion
(EUV irradiation, shock waves) are largely unknown, and should also be 
investigated.

\section*{Acknowledgments}
The authors are indebted to Dr. E. Halbedel, who provided us with  
unpublished photometric data on HD~53367, obtained at Corralitos Observatory. 
We are also very grateful to Dr. Luo Shao-Guang for sending us his 
100~$\mu$m optical depth scans of the CMa OB1 region.  We also are grateful 
to Dr. T.P. Ray for supplying additional
information on his spectroscopic observations and to Dr. R.D. Oudmaijer for 
providing us with electronic versions of his spectra of HD~52721 and HD~53367.  
Thanks are also due to the technical crews for support during the observations 
at ESO and the SAO.  This research has made use of the Simbad data base, 
operated at CDS, Strasbourg, France.

\bsp
\label{lastpage}
\end{document}